\documentclass[11pt,nofloats]{article}
\usepackage{graphicx}% Include figure file

\usepackage{setspace}
\usepackage[greek,english]{babel}

\begin{document}
\latintext
%\doublespacing
\centerline{\Large \bf Founding quantum theory on the basis of consciousness}
\vskip 0.2 in
\centerline{ Efstratios Manousakis}
\centerline{Department of Physics, Florida State University,}
\centerline{Tallahassee, Florida, 32306-4350, U.S.A and}
\centerline{Department of Physics, University of Athens, 
Athens, 15784, Greece.}
\centerline{email: stratos@martech.fsu.edu}
%\vskip 0.6 in
\vskip 0.2 in
\centerline{To be published in {\bf Foundations of Physics}}
%\centerline{Submitted: September 2, 2005}
%\centerline{Accepted: December 22, 2005}
\centerline{Publication date: June 6, 2006 (Found. Phys. {\bf 36} (6))}
\centerline{Published on line: DOI:  10.1007/s10701-006-9049-9}
\centerline{http://dx/doi.org/10.1007/s10701-006-9049-9}
%\vskip 0.1 in
 
In the present work, quantum theory  is founded on the framework of
consciousness, in contrast to earlier suggestions that consciousness might 
be understood starting from quantum theory. 
The notion of streams of consciousness, usually restricted to conscious 
beings, is extended 
to the notion of a Universal/Global stream of conscious flow of ordered events.
The streams of conscious events which we experience constitute 
sub-streams of the Universal
stream.  Our postulated ontological character 
of consciousness also consists of an {\it operator} which acts on a 
{\it state of  potential consciousness} to create or modify the 
likelihoods for later 
events to occur and become part of the Universal conscious flow.
A generalized process of measurement-perception 
is introduced, where the operation of consciousness brings into 
existence, from a state of potentiality, the event in consciousness.
This is mathematically represented  by (a) an operator acting on the 
state of potential-consciousness 
before an actual  event arises in consciousness and 
(b)  the reflecting of the result of this operation back onto the state of 
potential-consciousness for comparison in order for the event 
to arise in consciousness. 
Beginning from our postulated ontology  that consciousness is primary and
from the most elementary conscious contents, such as perception of periodic 
change and motion, quantum theory follows naturally as
the description of the conscious experience. 
\vskip 0.3 in
{ KEY WORDS:} Consciousness; quantum theory; measurement, EPR paradox.
\latintext
%\tableofcontents

\section{\large INTRODUCTION}
\label{introduction}

Quantum theory is plagued with conceptual difficulties such as the 
question of the so-called wave-function 
collapse in the measurement process\cite{wheeler}, and the well-known 
paradoxes\cite{wheeler,EPR,cat,vandermerwe}. 
While the literature addressing  various issues of the
interpretations of quantum 
mechanics\cite{heisenberg,bohr,bohm1,everett,wheeler,ballentine} is rather
rich\cite{vandermerwe,vandermerwe2,bell,stapp,stapp2},  in this paper 
we will limit ourselves to mention those more directly related to the
point of view elaborated here. Independently 
Schr\"odinger\cite{schrodinger,schrodinger2},
Wigner\cite{wigner}, Von Neumann\cite{von-neumann},
 London and Bauer\cite{london}, and Pauli\cite{pauli} 
have considered the possibility that consciousness plays a fundamental role 
in the interpretation of quantum mechanics, especially in the question of 
measurement. 

While in the so-called Copenhagen interpretation\cite{bohr} as well as 
in other interpretations the  observer enters in a
fundamental manner, ultimately, the nature of the observer is objectified
and is considered to be the same as the instrument which carries 
out the measurement. On the other hand, the paradoxes
and the problems associated with the process of the wave function 
collapse and the process of measurement have led many of the 
pioneers of quantum theory to examine the possible role of consciousness
in the process of measurement. For example, Schr\"odinger 
in his ``Tarner Lectures'' at the Trinity College (see ``Mind and 
Matter'' second part in Ref.\cite{schrodinger}, Chapters 1-4) 
discusses the fundamental
role of consciousness and the role of the process of objectivation in the 
description of nature. Von Neumann\cite{von-neumann}, 
using projection  operators
and density matrices as tools to describe the apparent statistical 
character of measurement, was able to show that 
the assumed boundary separating the observing instrument and the so-called 
observed object can be arbitrarily shifted and, therefore, ultimately 
the observer becomes the ``abstract ego'' (in Von Neumann's terms) 
of the observer. 
Similarly, London and Bauer\cite{london} followed a similar scheme 
to conclude that it is the ``creative action of consciousness'' 
by which ``the observer establishes his own framework of objectivity and
acquires a new piece of information about the object in question''.
According to London and Bauer\cite{london}
this leads to the collapse, namely to the choice of a particular 
state  from a linear combination
of the correlated states describing the combined system, i.e., the instrument 
and the observed object.
Similarly Wigner\cite{wigner} proposes that inanimate or unconscious 
matter evolves deterministically according to the quantum mechanical
evolution operator and when consciousness operates on inanimate matter,
the result is the familiar projection process of measurement.
Pauli and Jung\cite{pauli} through several letters and communications 
have discussed a parallelism between the operation of consciousness 
and quantum theory. 
Some of the philosophical ideas, from which we use in the present 
paper, can be also found in the books of 
Schr\"odinger\cite{schrodinger,schrodinger2}
and they are extensively discussed in the next section.

In this paper  quantum theory (more generally, the description 
of nature) is founded on the framework of the operation 
and on the primary ontological character of 
consciousness, rather than founding consciousness on the laws of
physics\cite{stapp,stapp2,penrose,nanopoulos}.
It is discussed that quantum theory follows naturally by starting 
from how consciousness operates upon a state of {\it potential consciousness}
and more  generally how it relates to the emergence or manifestation
and our experience of
matter. In addition, it is argued that 
the problem of measurement
and the paradoxes of quantum theory  arise due to
our poor understanding of the nature and the operation of consciousness.

\section{\large CONSCIOUSNESS AND ITS OPERATION}
\label{consciousness}

We begin  by introducing our concepts 
such as, streams of consciousness (particular and Universal), 
potential consciousness, and operation of consciousness which are necessary
in order to discuss our ontological proposal.

\subsection{Streams and Sub-Streams of Consciousness}

The word  consciousness usually means 
``experienced awareness''.  A
person is  ``conscious'' or  ``has'' consciousness if  he is  
experiencing a ``flow'' of conscious events.  The stream of consciousness 
consists of the conscious  events  that constitute  this  stream.   
The  order or  the
sequence of these events gives rise to a temporal order which together
with  the experience  of  temporal continuity\cite{poppel}, as we will 
discuss later, introduce  the
concept of  continuous time  used in physics  to describe the  laws of
nature. The contents of the stream, i.e., the conscious events, 
act to modify the
tendencies for later events to enter the stream. 
 The {\it subject}  is either the holder or  the experiencer 
of this ``flow''  of events  or just  this  ``flow'' of  entangled events.

First, we conjecture that all human beings and the other living organisms
have their own  streams of consciousness. In order to
gain an understanding of all of these related streams of consciousness
together,  and  what  precedes  our  human thoughts,  and  binds  them
together,  we postulate the existence  of the Universal/Global  stream  
of consciousness, as the primary reality that contains all of our 
individual streams, (which are sub-streams of the Universal conscious 
flow of events) and also conscious events
that are not members of any  human stream, but are like certain of our
conscious events (to be clarified in later subsections). 
Note  that the  set  of   conscious events  in
consciousness must include all those that anyone has ever had, and for
any  personal stream  of  consciousness, all  the events that  have
appeared in  that person's stream of  consciousness.

Even though all of our thoughts and experiences are in our stream of
consciousness, and have certain ``feel''  or {\it quale}, it is common
and useful to draw some distinctions between different kinds
of thoughts. While both are parts of our stream of consciousness,
experiences through the sensory apparatus, or sense data, are
distinguished from theoretical constructs where memory is a contributing
factor in causing these events. Therefore, when we distinguish between
``mind'' and ``matter'' we are referring to the previously 
mentioned distinction. ``Mind'' refers to our conscious experience
of the process of thought, whereas ``matter''  refers to 
the conscious experience of an imagined set of properties that are
imagined to exist ``when no one is looking''. A simple example of the
latter is the experience of a {\it surface}, which, under 
microscopic examination, the {\it imagined} notion of {\it surface} disappears
and is replaced by another {\it imagined} notion of an array of  
relatively widely spaced atoms or molecules. We are not denying
the existence of something that causes these experiences and it persists
when ``no one is looking''. Instead we are
only questioning its ``substantial nature'' and we postulate that 
these properties
{\it exist} in the Universal Mind, namely, they are parts of the 
Universal stream of consciousness. The emergence of ``matter'' is discussed
in Sec.~\ref{matter}.

\subsection{Intuition and State of Potential Consciousness}
We postulate that a fundamentally {\it new experience} or
an {\it insight} into a problem can occur through the {\it intuitive} 
character  of the mind (or consciousness). Aristotle speaks of intuition 
which is only  a
``possibility  of knowing without in any respect already possessing 
the knowledge to be acquired.''\cite{aristotleintuition} 
The conscious event of a new experience or of the insight, while after 
it happens it can be rationalized, prior to its occurrence it can not
be rationally inferred from the previous experiences alone\cite{kant}. After 
such an event becomes part of the individual's stream of consciousness,
it can be incorporated into reasoning by assigning the experienced conscious
meaning of it or qualia.  This can be demonstrated by the fact that we cannot 
use reason to directly explain the experience of color to someone 
born-blind or any experience to one who never had this particular experience.
Spinoza\cite{spinoza} clearly states that reasoned conviction is no help 
to intuitive  knowledge and Whitehead\cite{whitehead} accepts
that ``all knowledge is derived from and verified by, direct intuitive 
observation.'' Jung defines intuition as ``that psychological function which
transmits perception in an unconscious way.''\cite{jung0}

While the earlier sequence of events in a child's stream
of consciousness might be prerequisites for any new experience 
which causes the child's further development,  the new experience cannot
be grasped in terms of the previous experiences {\it alone}. 
The same argument seems to be valid in the process of the evolution
of species or of life in general. In order to describe such a process
of fundamentally new conscious events, we find it 
necessary to introduce the notion of {\it potential consciousness}.
Namely, in order to explain the manifestation of the experience and its
tendency for re-occurrence after its first manifestation, we  
conjecture the existence of the potentiality\cite{potentiality} 
for such a manifestation. When the
experience fades from the stream of consciousness, we conjecture that,
with respect to such an experienced quality, the state of consciousness
is transformed into its potential state again with a modified potential
for re-occurrence of this particular ``felt-quality'' in a future conscious 
event. 
  
In the case where the intuitive mind is constrained in such a way that
a fundamentally new experience (or intuition) is not allowed to occur, 
the state of 
potential consciousness is still necessary in order to describe what happens.
 One of the 
capabilities of human consciousness is
to {\it imagine} what  some future conscious  event might
be,  and   pre-ascribe  values   and  likelihoods  to   these  future
possibilities. In this case the sequence of events along with their character 
and conscious qualia form the basis on which the state  of potential  
consciousness is defined, namely, the potential for a particular future 
event to occur depends
on  the  contents of  consciousness  accumulated  during the  previous
events leading to the present.   

We also ascribe to the Universal consciousness a 
{\it Universal state of potential consciousness} out of which 
an event can arise in the Universal stream of consciousness.
Namely, just as the entire history of
events which form the  stream of consciousness  in an  individual and
have causal consequences on what can  happen in the present time or in
the  future time,  in the  same sense  the contents  of  the Universal
consciousness  and their temporal  order along  with the  operation of
consciousness could entail evolution, unless
a higher-level operation of consciousness intervenes. The operation of
consciousness and its hierarchical structure is discussed
in the following subsections.

Therefore, we  postulate the primary ontological status, the oneness, 
and the universality of  consciousness\cite{parmenides,parmenides2,
plato,vedanta,krishnamurti,spinoza,schrodinger}. 
The term ``oneness'' means that
there is only one stream of conscious flow with various sub-streams,
the individual streams of consciousness, such as those which we are 
experiencing as human beings, but they are all connected to one 
Universal conscious  flow. There is a hierarchical tree-like 
structure of this {\it ``branching''} of the
conscious flow as can be also evidenced within the human body. While
there is a vast number of sub-streams deriving from the Universal
consciousness, they all belong to the same single unbroken flow.
This separation  though between Universal  and individual  consciousness or
streams  of consciousness  is done  to facilitate the description of
our experience and there is no sharp boundary,  namely, the individual 
consciousness is a sub-stream of the Universal consciousness.

As it is discussed next, consciousness also {\it acts} 
on its state of {\it potential consciousness} in order to
cause an {\it alteration or change} of the state 
in order to cause an {\it experience}. Finally consciousness 
has also the faculty of the {\it experiencer}, namely, 
the felt qualities or qualia or the objects which appear through the
process of {\it perception or measurement}, a process which can take
place at the level of the particular sub-stream or at a lower- or
higher-level sub-stream of the Universal flow of consciousness. 
These notions are elaborated in the following three subsections. 

\subsection{The Operation of Consciousness}
\label{operation}

In our theory, consciousness also plays an active role 
and next we give an outline of the main functions of 
consciousness which are further discussed in the following sections.

As discussed earlier we postulate that the ontological character of 
consciousness is primary\cite{note5}.
The world ``outside'', which is perceived by consciousness as objective,
becomes actual through the operation of consciousness. Can one describe
the state of any part of nature {\it before} the operation of 
consciousness? Let
us call this state of nature before the operation of consciousness, 
``pro-nature''? 
Our language, our mathematics, our process
of thought, our experience, is based on concepts (or percepts) 
which are all contents  of consciousness. 
We will use the term conscious-concept or conscious-percept
to generally represent the conscious quale or felt experience (or 
simply percept).
Let us ask ourselves the question,  ``how else can we describe something
that is behind the  conscious perception?'', namely, by avoiding the
usage  of  any concepts or percepts,  because  they  are all products 
of  perception,
i.e. of  consciousness. Einstein taught us that even ``time'' and ``space''
 (which were once believed
to ``stand out there'' independently of us) exist, in a sense, dependent
on the process of their perception and measurement and they have no meaning
independently of that. In this paper, we take
a phenomenological position, that this ``pro-nature'' is also an aspect of
consciousness, namely, the best way to describe it and be free of contradiction
is to call it by its potential aspect and that aspect is ``potential to 
become conscious''. Therefore, the state of nature, before the application 
of consciousness, is a {\it state of  potential consciousness}.
In the previous subsection we discussed that from the 
state of potential consciousness and through the operation of 
consciousness on the state of potential consciousness, 
an actual event arises, and enters the 
 particular or Universal stream of conscious flow. 

Consciousness can be realized as operation and as  
experience\cite{lonergan} through perception as follows. 
First in order to describe the operation of consciousness at a {\it particular 
level} of the Universal stream, we need to divide the complete hierarchical
set of operations in two sets as follows: 
(i) The operations which occur at a {\it particular level} which corresponds
to a particular sub-stream of consciousness and 
(ii) the operations which occur at all other levels.  Namely,

\begin{itemize}

\item (a) First, the operation of consciousness at all other
levels can be thought of as an operation which causes the
state of potential consciousness, which corresponds
to a particular sub-stream, to change or evolve. This change or
evolution of 
the state of potential consciousness and its relation to
the perception of time, which is complementary to the perception of change,
is discussed in Sec.~\ref{change}.

\item (b) The state of potential consciousness evolves
and remains in potentiality until
it is perceived or measured by consciousness'
appropriate ``instrument'' at the particular level or sub-stream 
of the conscious flow. 
When this happens, an event arises or is actualized 
in the particular sub-stream of 
consciousness from the state of potential consciousness. 

\end{itemize}

{\it Perception} or an event in consciousness can only be 
{\it actualized} only if 
consciousness {\it operationally projects or measures} the 
experience or event  as follows:

\begin{itemize}

\item Consciousness asks a question (inquiry) or perceives a {\it change or 
alteration} in its  state of potential consciousness $|\psi_i\rangle$,  
by acting on the $|\psi_i\rangle$. The
result of this operation, i.e., $|\psi_{i+1}\rangle= \hat O 
| \psi_i\rangle$ (here $\hat O$ represents the action of consciousness, 
through an operational question which in general causes a change), 
is evaluated  by comparing the changed
state of potential consciousness, i.e., $|\psi_{i+1} \rangle$ with 
the previous state of potential consciousness $|\psi_i\rangle$.

\item This  process of {\it projection or objectivation} creates 
an actual event in 
consciousness. Namely the event is manifested or it becomes a 
phenomenon\cite{note1} or an object in consciousness through 
such an operation of  consciousness.

\end{itemize}

Therefore, we postulate a
sequence of pairs $\{ |\psi_i\rangle, C_i \}$ consisting of a sequence of
conscious events $C_i$ during each of which consciousness 
operates upon and changes the state of 
potential consciousness $|\psi_i \rangle$. The
state $|\psi_i\rangle$ constitutes  a set of potentialities out of which 
the next conscious event $C_{i+1}$ arise. Namely, the activities of 
our body/brain are not the
causes of conscious events, they are consequences of conscious events.
%As discussed earlier, human streams of consciousness are sub-streams of the 
%Universal stream of consciousness.

Let us give a simple example of the operation of consciousness.
The subjective experience of the sweet taste of a fruit. The fruit is
not sweet unless it is tasted, namely, there is no sweet taste attached to the 
molecules of the fruit. Sweetness is a subjective experience;
it is not a property of any interaction whatsoever of the ingredients of the
fruit with our mouth or with our nervous system. The result of this interaction
is only a network of electromagnetic currents from the neurons of the 
body/brain. The same is true with any (subjective) experience. The word 
subjective is placed inside parentheses because all experience is subjective
in the sense that only when it leads to such subjective qualia, it is
experienced. It appears through the process of objectivation\cite{note2} 
that while ``our hands'' 
(including man-made instruments, our eyes, the nerves carrying the signal 
to the cortex, etc) act on the state of ``nature'', what we actually ``see'',
is  the conscious experience, the event in consciousness and this is 
what we must describe. 

The question, ``is this fruit sweet between observations of 
its sweetness?'', is a meaningless question. The correct question to ask is:
``Is this fruit potentially sweet?'' or even better: ``What is the likelihood 
for this fruit to be found sweet when it is tasted?''. Clearly, 
the experience brings into existence, in consciousness, the taste from 
potentiality\cite{potentiality}.

Therefore, in general, when an event occurs,
an observable takes a definite value in consciousness from 
the state of the potentially conscious.
The value has only meaning with respect to a measure, an ideal, an observable
quality (in the example given before the measure is the taste and its 
value or  its quality that of sweet) acquired by the 
operation of consciousness. 
If consciousness does not operate 
because the conscious {\it attention} is not there (for example someone's 
attention is not
in what he is eating but elsewhere), there is
no event in ``one's  consciousness'' (in the particular sub-stream). 
When consciousness operates through one's
attention, only then  the operated state of potential consciousness 
is compared to consciousness' previous state. This is the process by which  
an event arises in consciousness. 

Next, we  would like to give  an example to  clarify why consciousness
needs to compare the  state obtained after the operational application
of the measure  or a question on the  state of potential consciousness
to its own previous state of potential consciousness.  Suppose that we
enter a  room for the first  time where an event took place before our
entrance; we may not  be able to figure out what the event
was;  however, a  person who  is ``aware''  of the  state of  the room
before  the  event is  able  to  find out  what  happened,  by a  mere
comparison of  the state of the  room before and after  the event.  In
the  same  way, consciousness  is  aware  of  its state  of  potential
consciousness before  the specific  observation, which takes  place by
altering  the   state  of  consciousness  by  the   operation  of  the
examination  of its  state operationally  and then  comparing  the new
state  with   the  old  state  of   potential  consciousness.  Namely,
consciousness only  perceives {\it change}  and this is  documented in
visual perception  experiments (see e.g.,  Ref. \cite{hubel}).  Namely,
if the retina  remains fixed relative to the image  there is no visual
perception by  the striate cortex neurons.  The constant microsaccadic
motion of  the retina  allows us  to see images,  and the  image fades
quickly if this motion does not occur\cite{hubel,vision}. 

Within this model how does one get  an experience of the room at all, if
one does not  have a prior experience of the room? However, the  
experience of  the  room itself is  a
change in the state  of one's own retina and post-retinal visual 
system and this is  why the room can be perceived.
In order  for that to  occur, the retina  as a
whole needs to move through  micro-saccadic motion; if the eye remains
constant relative  to the image, there  is no perception\cite{hubel,vision}.  
First at the retina level the photons are {\it perceived} through
the operation of consciouness on the receptor cells.
At  a higher-level in  the hierarchy  of the conscious
flow, the person still needs to perceive (operate  or cause a conscious 
event in ``his'' sub-stream of consciousness by activating the 
corresponding neural correlate to the person's consciousness) 
the image of the state prepared by the post-retinal visual system.  
Therefore consciousness  only  perceives  changes  by  comparing the state of
potential consciousness to the state obtained after 
the operation of the attention of consciousness.

Another related question which can arise here is the following:
It appears that related brain activity generally precedes the occurrence
of a perception which might make it difficult to accept the
idea that consciousness is primary. Does a related brain
activity precede the occurrence of the experience of an image? It is
true that  some brain activity  precedes the perception of  the image,
but is  that activity  the perception of  the image itself?  Perception comes
into being  through the intervention of consciousness,  and without it
there is  no perception. One can  be near an  image but he may  not be
seeing it. The retina may be ``seeing'' it but what we call the ``person''
(a higher-level conscious operator) does  not see it. We can imagine a
person with damaged striate cortex  with perfect eyes; he will not see
the image which his retina ``sees''. Therefore, the brain activity of the
pre-cortex visual system  still takes place with no  perception of the
image  by  the  person.  This  activity might  be  confused  with  the
perception  itself of  a more  processed image.  This argument  can go
deeper  in  the brain  until  we encounter  Von  Neumann's  notion of  the
``abstract ego''. The brain activity, which seems to precede the perception,
 occurs as any particular conscious sub-stream, corresponding 
to the various brain parts, becomes ``conscious'' (by means of a 
conscious event
entering the Universal stream through this sub-stream) of some precursor
of  the higher level perception,   in   the   vast tree-like
hierarchical  structure of flow of consciousness. 

These are simple examples from our everyday experience which are only given to 
demonstrate the process of how consciousness brings about events.
As we have already discussed, this does not imply that 
we should require the presence of a body of a ``sentient being'' 
for something to come into existence, because an event can arise in the
Universal stream of consciousness. The observed universe is the body of the
Universal consciousness! Many objections\cite{penrose} to 
previous attempts to interpret quantum theory using consciousness boil
down to the requirement that consciousness is owned by the bodies 
of sentient beings. 

\subsection{The Emergence of Matter}
\label{matter}

We have discussed the  state of potential consciousness and the operation
of consciousness on that state  which produces an event which is added
to the  stream of  consciousness.  These operations  can occur  at any
level  in  the stream  of  consciousness,  such  as at  an  individual
sub-stream  of  consciousness  or at  the   Universal  stream   of
consciousness.

What is the ontological status  of space-time and quantum fields?  
In this paper an attempt is made to construct an  ontology 
based fundamentally on consciousness. Then, we  take  both the  quantum
fields   and   space-time   structure,   that  gives   the   space-time
relationships between them,  to be basic elements in  what we call
the  Universal Mind or  Universal Consciousness.  Therefore space-time 
and quantum fields are experiences or contents of the universal stream
of consciousness to which our experiences all belong.  As we  
will show in Sec.~\ref{change}, the parameter of time used  in physics 
is  related  to the order of occurrence of these conscious
events which  take  place in  the  Universal  stream of consciousness. 
In Sec.~\ref{motion} we show that, 
the perception  of space and motion 
are also based on conscious events which, along with their associated felt
qualities, enter the Universal stream of consciousness.

The emergence  of
matter out of  the operation of consciousness occurs  at the Universal
stream of  consciousness, therefore, these  events seem to us,  to the
individual  stream  of consciousness,  far  more  stable, long  lived,
persistent,  namely, they  seem to exist ``when no one  is looking''. 
We generally  postulate that
when any new event occurs  at any level in the Universal stream
of consciousness,  it changes the state of  potential consciousness, and,
therefore,  it can  have observable  effects at,  or effects  that will
influence, any sub-stream.

The potential consciousness and the actor or  the operation of
consciousness are  primary.  The stream  of conscious events, which 
as  they occur
modify the  state of  potential consciousness for  later events  to be
added to  the stream of conscious events,  are emergent.  Consciousness
as an actor or an operator is beyond time.  
The Big-Bang itself is an event in the Universal stream  of consciousness.
However,  the Universal potential
consciousness and the operational consciousness are always present 
and the Big-Bang, as well as all other events, are manifestations of
the Universal potential consciousness.

\subsection{The Emergence of the Brain}
\label{brains}
 How does  Mind (consciousness)  use the  already  realized  events  
in consciousness  to
 allow the  manifestation of more complex perception or
concepts (which are  also  events)?  First,  these  rather
simple events discussed in the previous subsection enter 
the Universal stream, and then the emerging 
complexity  is the manifestation
of  higher-level  conscious   events  in  the  Universal  stream  of
consciousness.   Therefore  simple  events  flow  into  more  complex
structures  to represent  these higher-level  conscious  qualities or
concepts.  We can also represent life by this flow, and, at some level
of manifestation of the conscious potential, the cell emerges and 
then the brain/body emerges which are 
manifestations of  higher-level  conscious operations.
The little  streams  flow into  or  merge into  greater
streams to create a larger flow  or higher-level streams and these
higher-level streams merge  into higher-yet-level streams  and so on.
At some level the  neurons or neural networks or other structures
in the brain emerge as manifestations of concepts, feelings, memories
and so on.  Therefore, we conclude that any conscious quality, when it
becomes  manifest, has  a counterpart  in the  brain and in  an individual
stream of consciousness (which is also part of the Universal stream). 
The neurons or other central nervous system
structures are  the neural correlates or manifestations  of the
conscious concepts (or percepts)\cite{eccles}.

How do neural correlates represent concepts?
For the case of a brain, we postulate the restriction that
the action of the conscious concept on the potential consciousness
must be  to restrict the potential consciousness to  one in which
the probability for a particular concept to hold is unity (certainty): the
neural  correlate of the  conscious concept  must be  actualized.  All
brain  activities  incompatible with  the  conscious-concept and its
neural-correlate must be projected out.  In order for this to make sense a
correlation must  hold between the  concept in consciousness  and some
component part of the state  of potential  consciousness 
(as it will become clear, this is the  quantum state  of the
brain, defined  by tracing  over the other  degrees of  freedom except
those of the  brain). The pattern of brain activity,  can be a pattern
that has component parts scattered over the brain. As we discussed the
allowed set of  concepts in consciousness must include  all those that
anyone has ever had, and for any personal stream of consciousness, all
the  concepts   that  have  appeared   in  that  person's   stream  of
consciousness,  in association  with its  neural correlate; the  
concept will always have  its neural  correlates distributed over the brain, 
and the concept will be  able to actualize the corresponding neural 
correlates.  That  is, the binding
problem
is solved by postulating  that the concept causes the collapse
which actualizes the connection between the concept and the neural-correlates.

As we discussed we have postulated the existence of the state of the 
Universal potential consciousness.
The  particular brain contains  the neural  correlates  
of the concepts or qualities of the particular stream of consciousness.   
Because of the existence of the brain with its neural correlates
it makes sense to consider the personal
potential  consciousness,  which means  a  state  where we  pre-ascribe
likelihoods only to concepts which  already have a neural correlate in
the particular  brain. It is possible for a particular stream of 
consciousness to come ``in contact'' with the Universal consciousness.   
When this ``contact'' is 
established an  insight comes forth on the particular
stream of consciousness. For this to occur,  
the operator  which carries  out measurements in  the brain  
has to be  suspended.  This operator is made out of the old contents 
and it is acting  on the  already  existing neural  correlates.  
The process  of thought which  is a  process of  measurement in  the 
brain should momentarily halt  in  order for this contact 
with  the Universal state  of potential  consciousness to become  
possible.  The reason  is that  the  action of  measurement  
creates decoherence  and
collapses the  state of local potential consciousness  to a particular
concept with  an already  existing neural correlate.   Therefore, when
these local  measurements cease,  the state of  potential consciousness
becomes  coherently entangled  with  the  Universal  state  of
potential  consciousness.   When  this  Global   state  of  potential
consciousness is  established and all  the brain activities  cease, it
becomes possible to establish a correlation between a new concept from
the  Universal mind  and  a new  neural  correlate in  the brain and, 
thus, a new event enters the personal stream of consciousness.   We
postulate that this is the process of a new perception, the process of
creation, the  process of  evolution, the process  of the growth  of a
child, and the process of acquiring an insight.

\section{\large MATHEMATICAL DESCRIPTION}
\label{mathematical}

In Sec.~\ref{appendix} (appendix) we present as example where we use earlier 
contents of our stream of
consciousness (the real numbers) to project a new concept (the solution
to the equation $x^2+1=0$) onto  a basis formed by the old concepts
(the real numbers). In  the same  sense,  in  an  experimental situation  to
measure the position of a microscopic particle, we begin with the concept of
position which  is a macroscopic experience, and  we build instruments
appropriate to measure or to project this content of
our streams of consciousness.  This macroscopic experience
of space is a content of our streams of  
consciousness created from direct macroscopic experiences,
events which enter the stream of our consciousness by 
interacting with macroscopic objects.  We cannot use a real
conscious  being   to  interact   {\it directly}  with the microscopic  
world   and to  make measurements  in the way our brain does (it is actually
our human consciousness which does it through the brain)
 by direct perception as described in 
Sec.~\ref{brains}.   Instead, we construct  instruments to
measure quantities  based on our known concepts  and therefore 
they do not have the capability to measure an unknown concept to us. This is a
process of projection  and this process, as was demonstrated by the example,
can be described by writing the potential outcome of the 
Newton-Raphson operation as a linear combination of pre-ascribed
likelihoods for events, which correspond to known concepts, to occur. 

The example in the appendix  demonstrates that we can use
linear spaces  and operators to describe  a situation in  which 
we consider a new 
realm where perception of new concepts is required in order to be
able to build a rational description of our experiences there. 
However, due to restrictions of our own stream of consciousness,
such fundamentally new contents are not allowed to enter the stream.  
Therefore, in order to express the potential outcome of a
conscious operation (or measurement), our consciousness
uses as reference pointers the {\it old} contents 
of our  consciousness which entered the stream of our consciousness
as a result of  {\it earlier experiences} (or earlier conscious events
which correspond to definite neural correlates).  
The result of such a restricted operational observation 
is a random one from a pool of potential outcomes, which obey a well-defined 
distribution, if the question is repeated many times. 
 
The process of the operation  of consciousness can be formulated
mathematically, in order to describe the perception of matter.
As in the example of the appendix, we will make use of
a Hilbert space and operators acting inside this space to describe
functions of our consciousness and potential consciousness; the 
role played by the eigenstates and eigenvalues of such operators was also
demonstrated with the example. A more general mathematical description
is as follows: 

\begin{itemize}

\item We  begin from 
 the  experienced dualism  between consciousness (subject)
and  object (any experience  in consciousness). Note, however, that
both subject and the object (as experienced quality of the actual event which
enters the stream of consciousness) are aspects of consciousness.
The state of potential consciousness will  be
represented by a vector   in Hilbert space. Using the Dirac notation, 
we can write this state vector as $|\psi \rangle $ which is a linear 
combination of the 
basis vectors  $| i  \rangle $, with $i=1,2,...,N$, namely, as follows
\begin{eqnarray}
   |   \psi    \rangle   =
\sum_{i=1}^N   \psi_i  |   i   \rangle.
\label{basis}
\end{eqnarray}
The  vectors  $|  i  \rangle  $  give all possible states of
consciousness (states describing potential events)  for  the particular
observable (concept) in  question. In the case of a particular conscious
stream of a person, these concepts are also associated to specific neural
correlates scattered over the brain.  All the  $N$ vectors together form  
a complete
basis  set of  states,  namely, they  cover  all potential  outcomes.
In general, however, depending  on  the phenomenon  which  we  need  to
describe, $N$  can be  finite or infinite.  In addition,  the discrete
variable  $i$,  labeling  the  basis  elements, can  be  a  continuous
variable;  in this  case the  summation in  Eq.~\ref{basis}  should be
replaced by an integration.  The above linear combination implies that
the  observable is  not in  any of  the potential  states.   Unless an
observation takes  place, all we can say  is that there is  a state of
potentialities. This is so, not because we don't know what the actual
value of the observable is, it is so because there is no value in 
consciousness. What is the taste of a cake before tasting it?  Obviously,
this question is meaningless, the right question is: what is the potential
taste of the cake before tasting it?
\end{itemize}

The state vector which is represented as a linear combination of 
potential experiences represents the state of potential consciousness
not experiences in consciousness. It is through the operation of 
consciousness that one of the potential experiences can be
materialized. Because of the
{\it potential} nature of the state (i.e, that, which the state 
describes, is not actual yet before the
experience) it is written as a {\it mixture of possibilities}. Each possibility
is fundamentally distinct from any other. The result of the experience
while unique, prior to the experience itself (when it is in potentia),
should be written in such a way that it is a mixture or a sum of 
probability {\it amplitudes} for each one to occur as opposed to
just probabilities. The reason is that we need to end up with probabilities
after the experience not prior to the experience. This is so in order 
to allow for the operation of consciousness to take 
place and then carry out the measurement of the experience
by comparing the previous state of potential consciousness
with the one after the operation in order to have an event.
After this action
we end up with real events with a probability given by the square 
of the coefficient
in the linear combination multiplying the particular state that
becomes manifest.

\begin{itemize}

\item Because consciousness needs  to  carry out measurements 
(operations) inside this space, to make an   event happen, this vector
space should have the property  of finite measure and, as a result of
these  requirements, it is a  Hilbert space.   In such  a  space, a
measure of  the ``overlap''   between two states  $|\psi \rangle$
and $|\phi \rangle$   is measured by the  scalar product between
the  two  vectors  representing  the  two  states,  namely, 
\begin{eqnarray}  
I = \langle  \psi | \phi \rangle = \sum_i \psi^*_i \phi_i.  
\end{eqnarray}   
The overlap of any state
to itself,  which is the  square of the  length of  the  vector, is
normalized to unity, i.e.,  $\langle \psi | \psi \rangle=1$ and
this is  possible when working    in a Hilbert  space. Depending on
the nature of  the observed phenomenon,   $\psi_i$  can be real or
complex  numbers,   or  other   mathematical  objects      such  as
multi-component vectors.       Then, consciousness operates,
by means of a linear operator acting on this state vector representing
potential  consciousness.  
An event in consciousness is a change and this change,
in general, is an operation acting on a previous state.  How
does consciousness  measures this  potential change?  If  attention is
absent there will  be no conscious event.  For such  an event to occur
in consciousness, consciousness  has to compare this state  to its own
previous  state   for  the  event  to  occur.   

\end{itemize}

The  state $|\psi \rangle$  represents the  
state of  potential consciousness,
which is not {\it realized} yet. It  can potentially lead to a real event or
experience through  the application  of consciousness or  attention of
consciousness.  The so-called  real  or physical  event  is a 
conscious quality (or quale) in the Universal stream,  which
consciousness  projects  ``out  there''.   There  is  no  difference  or
separation between the  qualia and the real,
the  physical. This notion was discussed in Sec.~\ref{operation}.  
 The   physical     is    an     experience  in the Universal
 stream of consciousness. In the particular or individual stream another
corresponding event enters, the one actualized in its central 
nervous system, when the observation by that person occurs.
Therefore,  the  state  is  defined  over  potentially
physical  events,  or  potentially  conscious events.  

The ``physical'' or the mental event, or simply
event,  can be  mathematically broken  down into  a two  step  process, (a) 
 the action or an operation which applies 
a concept on the state of potential consciousness and transform the state
of potential consciousness (by activating the corresponding object 
(or ``material'') correlate)
to a state representing the concept alone, 
$|\phi \rangle = \hat O | \psi \rangle$ ( $\hat O$ is  the  
operator  representing  a  particular  action  of  consciousness) 
(b) the overlap of the 
changed state (after  the  operation)  to  the state of potential consciousness
prior  to  the operation, i.e., 
$ M = \langle \psi | \phi \rangle$, corresponds to the conscious value or
the quality of the applied concept. 
After this process, the potential 
becomes real, namely, it appears in consciousness, or equivalently,
it {\it is} a physical event and it activates the objective-correlate.
 
In the example given in the appendix, how does consciousness  
find the solution to an equation?
Using the language of the operational consciousness
this can be formulated as follows. The Newton-Raphson operator $\hat O$ 
creates the potential solution of the equation.  When 
the state $|x_{n+1} \rangle =
\hat O | x_n \rangle  $ and the previous state $| x_{n} \rangle$ have 
large overlap we take it that the solution is found (or ``observed''). 
This is how we decide that we have found the solution,
namely when $\langle x_n | \hat O | x_n \rangle = 1$,
within a resolution defined by our computer precision.
In order to make sure that we found a solution independent of the
initial condition, we may start from a state
$| \psi_0 \rangle = \sum_{l=1}^M | x^l_0 \rangle$ and after
application of the operator $\hat O$ several times, we stop when
the overlap $\langle \psi_n | \hat O | \psi_n \rangle$ is maximum
(or unity, if we keep normalizing the states $| \psi_n \rangle$).

\begin{itemize}

\item Therefore, every creative action of 
consciousness  can be  mathematically represented by 
an operator $\hat O$  applying an idea or a concept to the state 
of {\it potential consciousness} $|\psi \rangle$. 
This causes a change of the state of potential consciousness. 

\item This changed state of potential consciousness $|\phi \rangle = 
\hat O | \psi \rangle$, due to the creative 
operation of consciousness,   remains in a state of 
potentiality until it is perceived or measured by consciousness' 
appropriate instrument, again through the operation of consciousness. 
When this happens the event arises in consciousness 
from the state of potential consciousness.

\item This new action of consciousness which causes
the {\it observation-perception-measurement} 
is completed by the comparison of the two states, namely  the one 
before the operation of consciousness, i.e., $| \psi\rangle$, 
with the one after the
operation of consciousness, i.e., $\hat O 
| \psi \rangle$, which is taken to be the scalar product between the two 
states:
\begin{eqnarray}
M  = \langle \psi | \hat O | \psi \rangle.
\end{eqnarray}
The result of this comparison is also the observed result of the 
measurement as was discussed in Secs.~\ref{operation} and \ref{mathematical}.
When this operator is used to represent a real (non-complex) 
physical observable, i.e., $M=M^*$, the operator $\hat O$ is a 
Hermitian operator. More information on the properties of a Hilbert space 
and of Hermitian operators acting in such a space can be found in 
Ref.~\cite{von-neumann}.

\item Each particular operation of consciousness, represented by an 
operator $\hat O$ that represents a particular observable or observing
operation, is characterized by {\it eigenvectors and eigenvalues} in the
Hilbert space, namely,
\begin{eqnarray}
\hat O | \lambda \rangle & = & \lambda | \lambda \rangle.
\end{eqnarray}
 The significance of the eigenvectors
of $\hat O$ is that these are the only states of potential-consciousness that
{\it do not change} by the particular act of consciousness, namely,  
through the 
application of the inquiry $\hat O$.
The result of the measurement (or the conscious quality) 
is the corresponding eigenvalue because the
projection of the result of the action, i.e.,  $\hat O | \lambda \rangle$
on the state $| \lambda \rangle$ itself, is the eigenvalue $\lambda$.
 %Therefore, they represent fundamental ``building blocks'' of consciousness. 
The eigenstates are the only states which represent a lasting experience
in consciousness through the perception which corresponds to the eigenvalue. 
This point is discussed by means of an example in the Sec.~\ref{appendix}.
Next we schematically discuss the main points.

\end{itemize}

Let us consider two such  operators, the operator  $\hat O$ and its 
eigenstates/eigenvalues as defined above, and the operator $\hat Q$  
with the following spectrum of eigenstates/eigenvalues:
\begin{eqnarray}
\hat Q | \mu  )   =  \mu | \mu ).
\end{eqnarray}
Since the eigenstates of each of these operators form what we
call a complete basis set of a Hilbert space, let us express any of 
the eigenstates of  the operator $\hat O$ in terms of eigenstates of 
the operator $\hat Q$, namely:
\begin{eqnarray}
| \lambda \rangle & = & \sum_{\mu} \psi_{\lambda}(\mu) | \mu ),\\
\psi_{\lambda}(\mu) & = & ( \mu | \lambda \rangle.
\label{mu}
\end{eqnarray}
The meaning of this expression is as follows. 
First let us suppose that the measurement of the observable 
(or question) represented by the operator $\hat O$ transforms the state 
of potential-consciousness to a particular eigenstate $|\lambda \rangle$.
The result of the measurement is the corresponding eigenvalue $\lambda$.
The next observation or question
to ask is represented by $\hat Q$, which may or may not be compatible with 
the previous observation $\hat O$. The result of a single observation 
corresponding to $\hat Q$ will transform the state of potential consciousness
to an eigenstate $|\mu )$ of $\hat Q$ corresponding to a definite conscious
quality characterized by the eigenvalue $\mu$. The result of a single 
observation/measurement will  bring about in consciousness only a single 
definite answer.
This answer must correspond to an eigenstate of the operator $\hat Q$ because
only the eigenstates of an operator are ``robust'' or ``lasting'' against 
the application of $\hat Q$.
As we have already mentioned, this is the reason why we use eigenstates to 
represent any particular realizable state of potential consciousness.
The particular state $|\mu )$ which would be brought about in consciousness
cannot be known (as discussed in Sec.~\ref{operation}), all that is known 
is that the previous state of potential consciousness is $|\lambda\rangle$.
This particular state that arises in consciousness is a choice that 
consciousness makes. 

In Sec.~\ref{appendix} we show that because of the 
limitation of consciousness' 
observing  instrument, the only way to possibly describe 
such an act of measurement is a distribution;
namely, any one particular state  is not a predictable 
outcome, whereas a particular distribution can be a predictable output 
of many measurements.

For the case of our example given above, this means that 
while the state of
potential consciousness is $|\lambda \rangle$, i.e., an eigenstate of the
observable represented by the operator $\hat O$,  consciousness carries out 
a measurement of an observable represented by the operation $\hat Q$.
A particular question can be the following $P_{\mu}:$ ``Is the state of 
potential-consciousness
the one corresponding to the eigenvalue $\mu$?'' This question is operationally
applied using the projection operator defined as follows:
\begin{eqnarray}
\hat P_{\mu} | \mu ' ) = \delta_{\mu \mu '} | \mu ' );
\end{eqnarray} 
i.e., such that the outcome of its operation on the state $|\mu ' )$ and then
projected back to itself (measured against itself) is given as
\begin{eqnarray}
( \mu ' | \hat P_{\mu} | \mu ' ) = \delta_{\mu\mu '}.
\end{eqnarray}
Namely, it is affirmative or negative depending on whether or not the state of 
potential consciousness agrees with that sought by means of the operational 
question $P_{\mu}$.

If the same question is applied on the state $|\lambda \rangle$ given 
by Eq.~\ref{mu}, we find
\begin{eqnarray}
\langle \lambda | \hat P_{\mu} | \lambda \rangle  = | \psi_{\lambda}(\mu)|^2,
\end{eqnarray}
namely, the outcome of the projection
would be the eigenstate $|\mu )$ with eigenvalue $\mu$ and with probability
$|\psi_{\lambda}(\mu)|^2$.  Therefore, we can represent the projection 
operator in Hilbert space as $\hat P_{\mu}  = | \mu  ) ( \mu |$.

\section{\large COMPLEMENTARITY OF CHANGE AND TIME}
\label{change}

Periodic change or fluctuation is a fundamental  element of consciousness. 
Consciousness perceives time only through the direct 
perception of change through an event; the value of the time interval 
between two 
successive events in consciousness is only found  by counting
how many revolutions of a given {\it periodic event} took place during 
these two events. Therefore the notion of time is related to the
sequential (ordered) events which allow counting, and the interval of time 
and change (in particular periodic change) are complementary elements
and they are not independent of each other.

There is physiological evidence suggesting the
{\it direct} perception of frequency. For example, we perceive the
frequency of sound directly as notes or pitch, without having to
perceive time and understand intellectually (after processing) 
that it is periodic. Another evidence of direct perception of 
frequency comes from the fact 
that color is perceived directly without the requirement that ``one's''
consciousness is aware of any co-experience of time whatsoever.
In addition, the retina receptor cells are highly sensitive and it has been
shown that they can observe a single photon\cite{receptors,visionRMP}.
Furthermore, in biological systems, receptors for what we refer to as 
time do not exist\cite{poppel}. On the contrary, there is
significant neuro-physiological evidence that
the perception of time takes place via coherent neuronal 
oscillations\cite{singer} which
bind successive events into perceptual units\cite{poppel}. 

Nature responds to frequency very directly, and some examples are
resonance, single photon absorption and in general absorption at 
definite frequency.  The timeless photon,
in addition to being a particle, can be thought of as
the carrier of the operation of consciousness on the 
state of potential consciousness together with the correct 
instrument. When the operated state of potential consciousness 
is measured against its own state before the operation, a definite frequency
is realized (or materialized). An instrument (such as the retina
receptor cells) is needed to materialize the operation of consciousness,
because matter is the necessary ``mirror'' to ``reflect'' (to actualize) 
the act of consciousness. At first glance it may appear that we have introduced
a duality by separating consciousness and matter. Matter, however, as discussed
in Sec.~\ref{matter} is manifested consciousness at another level, at the 
Universal stream of conscious events.

Let us try to discuss evolution (or change) quantitatively. 
In order to describe any perception of change, our imagination 
invents a parameter which we call {\it time}
which labels the various phases of change and we delude ourselves with 
the belief that such a parameter has independent existence
from consciousness; time is only a vehicle or a label used to facilitate the
description of change. Therefore, we imagine the state of potential 
consciousness $|\psi(t) \rangle$, as a function of $t$, labeling the time of 
potential observation.  We wish to discuss a periodic motion, so 
let us confine ourselves within a cycle of period $T$. For simplicity 
we will discretize time, namely, the states are labeled as 
$| \psi(t_i) \rangle$ where $t_1=0,t_2=\delta t,t_3=2 \delta t ,..., 
t_N=(N-1) \delta t$, with $N\delta t = T$. 
These time labels have been defined and 
measured in terms of another much faster periodic change which we call it
{\it a clock}. Let us assume that $\delta t$ corresponds to 
the ``time'' $T'$ of a single period of the fast periodic change 
of the clock, namely $t_i$, $i=1,2,..,N$, are the moments when the ``ticks'' 
of the clock occur.  
Let us define the {\it chronological operator} $\hat t$ and its eigenstates
\begin{eqnarray}
\hat t | \psi(t) \rangle = t | \psi(t) \rangle,
\end{eqnarray}
namely, we have assumed that the state of potential consciousness
is characterized by a definite measured time.  
 Notice, that we needed two periodic motions ``running'' in parallel 
in order to discuss the measurement of the period of the first in terms 
of the second (clock).
Namely, we are unable within a single event in consciousness
to know both the time and the frequency of the event. We have
already discussed that physiological
evidence given above, suggests that consciousness only experiences 
frequency not time as a fundamental conscious quality. 
Here we have put the cart before the horse by beginning from the 
imagined notion of time, which is only quantified through the periodic 
motion.  We will next define the eigenstates characterized by definite
period in terms of the states characterized by definite time.

Next, let us define the evolution operator or
time-displacement operator, the operator that causes the change of the
state of potential consciousness, namely
\begin{eqnarray}
\hat T | \psi_i \rangle & = &  | \psi_{i+1} \rangle,\hskip 0.2 in i=1,...,N-1,
\nonumber \\
\hat T | \psi_N \rangle & = &  | \psi_{1} \rangle,
\label{periodic}
\end{eqnarray}
where $|\psi_i \rangle = | \psi(t_i) \rangle $, and
the second equation above implies that because of the nature of the 
perception of the periodic change there will be no difference in the state
after time $t=N\delta t$, i.e., the period $T$ of the periodic change.

In the case of periodic change, such as described by 
Eq.~\ref{periodic}, all the eigenstates and eigenvalues of the operator
$\hat T$,  in terms of the eigenstates of the chronological operator,
are given as follows:
\begin{eqnarray}
\hat T | \omega_n \rangle & = & \tau_{\omega_n} | \omega_n \rangle,
 \hskip 0.2 in
\tau_{\omega_n}  =  e^{-i \omega_n \delta t}, \\
| \omega_n \rangle & = & {1 \over {\sqrt{N}}} \sum_{i=1}^{N} 
e^{i  \omega_n  t_i} | \psi_i \rangle,
\label{frequency}
\end{eqnarray}
where $\omega_n  = n ({{2 \pi} / {T}})$, and $n=1,2,...,N$.
Notice that the quantization of the levels of single periodic change 
is the same as that of the harmonic oscillator (here we have used 
{\it natural} units where  the so-called ``energy'' is the same as 
the frequency and this is discussed later).
This is so because we have not limited in any way the number of quanta 
(or atoms) which are observed.

The state which describes a periodic change is such that when the time 
displacement operator acts on it, it behaves as its eigenstate.
The significance of the eigenstates was discussed in Sec.~\ref{mathematical}.
Namely, they are the only states of potential consciousness which do
not change under the application of the inquiry, no matter how 
many times the inquiry (measurement) is applied. 
Therefore, the measurement in this case
introduces no frequency uncertainty of the state of definite frequency.
This state cannot be characterized by any definite value of time; for the 
change to be characterized by a definite frequency, an observation of 
regular periodic motion is required to continue forever.

Time $t=m \delta t$ ``elapses'' when the time-translation 
operator $\hat T$ acts $m$ consecutive times on the state, namely, the time
evolution of the state $|\omega_n\rangle $ is
\begin{eqnarray}
| \omega_n\rangle_t & = & \hat T^m | \omega_n \rangle  = 
e^{-i  \omega_n  m \delta t} | \omega_n \rangle \\
& = & e^{-i  \omega_n  t} | \omega_n \rangle.
\end{eqnarray}

Let us now consider the case where the change does not necessarily occur
at a single period but it is a mixture of periodic changes of various
characteristic frequencies. We begin from the frequency eigenstates 
Eq.~\ref{frequency} as the basis and let us define the time evolution 
of the state as
\begin{eqnarray}
| \psi(t) \rangle & = & \sum_n c_n | \omega_n \rangle_t \\
& = & \sum_n c_n e^{-i\omega_n t} | \omega_n \rangle,
\end{eqnarray}
where the sum is over all the eigenstates of the time translation operation
acting on the state of potential consciousness that characterizes the system.
This latter equation can also be written as follows
\begin{eqnarray}
|\psi(t) \rangle = e^{-i \hat \omega t} | \psi(0) \rangle,
\label{time-solution} \\
\hat \omega | \omega_n \rangle = \omega_n | \omega_n \rangle,
\end{eqnarray}
where $|\psi(0) \rangle$ is some reference state. Again the perception of
frequency is direct and so in our description of 
nature we need to start by  considering this perception
as one fundamental building block of consciousness and not 
the perception of time. 

Equivalently from Eq.~\ref{time-solution}, we can say that
this is the solution to the following differential equation
\begin{eqnarray}
\hat \omega | \psi(t) \rangle = i {\partial_t} | \psi(t) \rangle,
\label{schr-equation}
\end{eqnarray}
or equivalently 
\begin{eqnarray}
\hat \omega = i \partial_t.
\label{freqoperator}
\end{eqnarray}
We need to discuss why the above frequency operator characterizes the 
measurement of change. For simplicity let us go back to the discrete time
domain. If consciousness applies the operator $\hat \omega$ on the state of
potential consciousness we have
\begin{eqnarray}
  \hat \omega | \psi(t_i) \rangle  = 
{i \over {\delta t}} (| \psi(t_{i+1}) \rangle - | \psi(t_i) \rangle ),
\end{eqnarray}
which is (apart from the multiplicative factor $i/\delta t$) the {\it change} 
of the state of potential consciousness.
This change is evaluated by simply using as measure the instantaneous 
state of potential consciousness itself, i.e.,  by projecting
the change onto  $| \psi(t_i) \rangle$. This means that 
the expectation value $\langle \psi(t) | \hat \omega | \psi(t) \rangle$
is a measurement of the rate of change of potential consciousness. 

Using Eq.~\ref{freqoperator}  for the  frequency operator,
the following commutation relation between frequency and the 
chronological operator follows in a straightforward manner:
\begin{eqnarray}
[\hat t,\hat \omega] = i.
\label{freqcommute}
\end{eqnarray}
In addition, the well-known uncertainly relation follows, namely, 
\begin{eqnarray}
\Delta \omega \Delta t \ge 1. 
\label{frequncertainty}
\end{eqnarray}
The uncertainty relationship between frequency and time can be easily
understood as follows. Let us suppose that  a changing state of consciousness
is observed for a finite interval of time $\Delta t$.
This observation time interval is also the uncertainty in time, 
because there is no particular instant of time inside this interval
to choose as the instant at which the observed event happened.
Even if the event seems to be regular or periodic inside this
interval of time, there is an uncertainty as to what happens
outside this interval. In fact, as discussed, nothing happens outside 
this interval because there is no observation in consciousness and thus 
no event there, only a potentiality. 
If we calculate the Fourier spectrum of such a changing event, no matter how 
regularly it evolves inside this interval, (with nothing happening 
outside of this interval) we will find significant amplitudes 
for frequencies in an interval range greater than $1/\Delta t$.

\section{\large COMPLEMENTARITY OF MOTION AND SPACE}
\label{motion}

The next question which naturally arises is how consciousness perceives
motion. Motion is associated with change of relative position. 
Let us inquire how consciousness perceives motion of a point-like object.
In  translationally
invariant space, how does one know that motion occurs? There is no movement 
unless there is an observer and a change of the position of 
the object relative to that of the observer. 
Just as we did in the previous section, where we considered 
frequency (periodic change) and time as 
complementary observables, here, we will consider regular motion in space 
and spatial position as complementary observables in consciousness; 
namely,  one needs the other in order to be perceived in consciousness.
There is a tendency to think that space is out there ``standing'' even if
the perception of motion was not there. However, the very 
definition of space requires the pre-conception of motion and the
perception of space implies motion as a potential event. 
In the following section, we will discuss that motion is also a
particular form of change and therefore we will require a relationship between
frequency and wave-vector. For the case of motion, however, there is
the field of space which can be used by consciousness to express this 
particular form of change, namely, motion; hence, momentarily
time can be set aside. This point will become clear below.

Let us examine whether or not we can use the state of potential-consciousness 
representing position in space to understand the state of 
potential-consciousness representing motion.
If a particle is observed to be in a particular fixed position in space, 
for example ${\bf r}$, we will represent the state of 
potential-consciousness by a state vector
$|{\bf r} \rangle$. In order for this state to successfully represent the
state of definite position, the operation of observing the eigenstate of the 
position should leave this state unchanged. 

For the case of motion let us begin from these eigenstates 
of position and work in a bounded world with periodic boundary conditions.
We then define the displacement operator $\hat T_{\delta {\bf r}}$, 
that causes the motion in consciousness, as follows:
\begin{eqnarray}
\hat T_{\delta {\bf r}} | {\bf r} \rangle = | {\bf r}+\delta {\bf r} \rangle,
\end{eqnarray}
and if the position vector  lies outside of the boundary of space it is mapped
inside using the Born-Von Karman boundary conditions.
In order to simplify the discussion, let us consider a one
dimensional problem with periodic boundary conditions, namely
a problem on a circle of length $L$ with discrete positions 
$x_i = (i-1) \delta x$ labeled as $i=1,2,...,N$, and $N \delta x = L$. 
Then the space-displacement operator is
\begin{eqnarray}
\hat T_s | x_i \rangle & = & | x_{i+1} \rangle, \hskip 0.2 in i=1,...,(N-1),\\
\hat T_s | x_N \rangle & = & | x_1 \rangle.
\end{eqnarray}

In order for consciousness to perceive motion, we need to define a 
state of potential-consciousness  which, when the displacement
operator acts on it, does not change it. 
Mathematically, assuming that the eigenstates of the position operator
form all possible outcomes of a measurement of position, it is possible 
to write down all the eigenstates and eigenvalues of the operator
$\hat T_s$ in terms of the eigenstates of the position operation.
They are given as follows:
\begin{eqnarray}
\hat T_s | k \rangle & = & \lambda_k | k \rangle, \hskip 0.2 in
\lambda_k  =  e^{-i k \delta x},  \\
| k \rangle & = & {1 \over {\sqrt{N}}} \sum_{l=0}^{(N-1)} 
e^{i  k l} | x_l \rangle,\label{momentum}
\end{eqnarray}
where  $k  =  ({{2 \pi}/ L}) j$, with 
$j=0,1,2,...,(N-1)$ and the states $| x_l \rangle$ are the position 
eigenstates along the circle.
The state which describes a periodic motion is such that when the time 
displacement operator (representing consciousness) acts on it, it does not 
change no matter how many times consciousness applies the inquiry. 
Therefore, the measurement in this case
introduces no wave-number (k) uncertainty of the state of definite 
$k$. This state cannot be characterized by any definite value of position.

Since the basis $k$ forms a complete set, we can express the 
position basis as a linear combination, namely
\begin{eqnarray}
| x_i \rangle  & = & {1 \over {\sqrt{N}}} \sum_{j=0}^{(N-1)} 
e^{-i  k_j x_i} | k_j \rangle, \hskip 0.2 in  k_j  =  {{2 \pi} \over  L} j.
\end{eqnarray}
These equations can be generalized from a discrete one-dimensional index to a 
continuous three-dimensional one, in the usual way. Namely,
\begin{eqnarray}
\hat T_{\delta {\bf r}} | {\bf k} \rangle & = & \lambda_{\bf k}
| {\bf k} \rangle, \hskip 0.2 in
\lambda_{\bf k}  =  e^{-i {\bf k} \cdot \delta {\bf r}},  \\
| {\bf k} \rangle & = & {1 \over {\sqrt{V}}} \int d^3 r 
e^{i {\bf k} \cdot {\bf r}} | {\bf r} \rangle, 
\label{momentum2}
\end{eqnarray}
where ${\bf k}  = (k_x,k_y,k_z)$, and for periodic boundary 
conditions each of the components is given by
$k_w  =  {{2 \pi} / L_w} n_w$ ($w=x,y,z$), with $n_w$ taking integer values
and $L_w$ are the dimensions of the box of volume $V$ bounding the space.

The position eigenstate $| {\bf r}' \rangle$ can be reached from 
$|{\bf r}\rangle$
by acting with the space-displacement operator as follows
\begin{eqnarray}
| {\bf r}' \rangle & = &  e^{i  \hat {\bf k} \cdot ({\bf r}'-
{\bf r})} | {\bf r} \rangle, \\
\hat {\bf k} | {\bf k} \rangle & = & {\bf k} | {\bf k} \rangle.
\label{momentum-eigenstate}
\end{eqnarray}
It is straightforward to see from the last equation, that 
the operator $\hat {\bf k} = (\hat k_1, \hat k_2, \hat k_3)$ is given by
\begin{eqnarray}
\hat k_i = -i \partial_{x_i},
\label{waveoperator}
\end{eqnarray}
where $x_i$, $i=1,2,3$ are the three components of ${\bf r}$.
As was discussed in the case of the frequency operator,  in a similar way 
it can be shown that the above wave-vector operator characterizes the 
measurement of change through motion. 
Namely, when consciousness applies the operator $\hat {\bf k}$ on the state of
potential consciousness
apart from the multiplicative factor, the {\it change} 
of the state of potential consciousness is obtained.
This change is evaluated by projecting
the change onto the state itself. This means that 
the expectation value $\langle \psi | \hat {\bf k} | \psi \rangle$
is a measurement of the ``rate''of change of potential consciousness 
with respect to spatial variation.

Using Eq.~\ref{waveoperator} for the wave-vector operators, it can be shown
in a straightforward manner, that the 
following commutation relation between position and momentum operators,
\begin{eqnarray}
[\hat x_i,\hat k_j] = i \delta_{i,j},
\label{wavecommute}
\end{eqnarray}
as well as the following uncertainly relations,
\begin{eqnarray}
\Delta x_i \Delta k_i \ge 1, 
\label{waveuncertainty}
\end{eqnarray}
are valid. This uncertainty relationship can be easily understood
by means of a similar argument as one provided for the case of 
the frequency-time uncertainty (Eq.~\ref{frequncertainty}).

There is direct experimental evidence that consciousness perceives
directly states of well-defined 
wave-vector\cite{hubel,vision,direction-neurons}.
There is a large number of striate cortex neurons which only respond
to motion in a well-specified direction\cite{hubel,vision}.
In particular in Ref.\cite{direction-neurons} the analog of the
two slit interference
experiment is introduced for the visual perception of the mammalian
brain. The response of the cat striate cortex neuron to
a single line of light flashed alternatively at two parallel locations
separated by distance $d$ 
was recorded. The response of the direction-sensitive 
neuron of the striate cortex (area 17) was found to fit the form
\begin{eqnarray}
f(d) = \sin(k d + \delta) e^{-d/\xi},
\end{eqnarray} 
as a function of the distance $d$ and $k=2\pi /\lambda$, with $\lambda$
the distance where the optimum response is found. Therefore
the mammalian brain's study indicates that  consciousness is 
wave-length selective. Indirectly this fact was already known to us, 
because we can see definite colors. 

Our operations, applied on observation instruments of the outside world, 
must be the same operations which we apply inwardly, otherwise what 
makes us apply different operations for observing two parts of the 
same world? Because the boundary between inward and outward, between 
the observed and the observing instrument can be arbitrarily 
shifted\cite{von-neumann,london}.

Therefore, consciousness naturally understands motion as well as position
as fundamental elements of consciousness. 
However, as it was discussed, they are also complementary observables,
each having no independent existence from the other. 
The Newtonian conception of classical mechanics (the notion 
of the independent existence (absolute) of the framework of space and 
time and, as a result,
the notions of rates of change) through the successful application of its
laws to describing all macroscopic motion, has given a tremendous credit
to these notions as existing independently. However, as we have recognized 
now, this conception is not accurate. 

\section{\large EQUATIONS OF MOTION}
\label{equations}

\subsection{Non-Relativistic Quantum Mechanics}
Let us summarize what has been shown so far. First, the
expressions for the wave-vector (Eq.~\ref{waveoperator})
and frequency (Eq.~\ref{freqoperator}) operators are generally derived. 
Using these relations, the commutation relations between position and 
momentum operators given by Eq.~\ref{wavecommute} 
as well as the uncertainty relations Eq.~\ref{frequncertainty} and 
Eq.~\ref{waveuncertainty} follow in a straightforward manner.

In the case where the uniformity of space is broken, namely the
various positions of space are biased differently (for example by 
an external field), the situation
is somewhat different. Let us consider the discretized one-dimensional 
space (with periodic boundary conditions) in order to demonstrate what 
happens in this case. The basis state vectors
are $| x_i \rangle $, labeled by the discrete positions $x_i=(i-1)\delta x$
where $i=1,2,...,N$ and $N \delta x = L$. When the frequency operator
is applied on any such position eigenstate $| x_i \rangle$ we can write
the following general expression
\begin{eqnarray}
\hat \omega | x_i \rangle = \epsilon(x_i) | x_i \rangle - (t_1 |x_{i+1} \rangle
+ t^*_1 |x_{i-1} \rangle) - (t_2 |x_{i+2} \rangle
+ t^*_2 |x_{i-2} \rangle) - ....
\end{eqnarray}
Notice that the coefficients of $ | x_i+n \rangle$ and
$ | x_i-n \rangle$ must be complex conjugates because the operator
$ \hat \omega$ is
Hermitean as discussed. Note that in general the coefficients
$t_i$ could also depend on $x_i$ but we consider the simplest case.
Let us initially consider only ``nearest-neighbors hopping'', i.e., 
we neglect all the terms $t_n$ with $n>1$.
Furthermore we can choose an overall phase factor such that $t_1$ is
real. It is straighforward to show that in the continuum 
limit ($\delta x \to 0$) we obtain
\begin{eqnarray}
\hat \omega | x \rangle = \Bigl ( u(x)  -{ 1 \over {2 \mu}} {{d^2}
\over {dx^2}} \Bigr ) | x \rangle,
\end{eqnarray}
where $u(x) = \epsilon(x)- 2$ and $1/2\mu = t (\delta x)^2$. Note
that if we consider the terms proportional to $t_n$ with $n>1$, they
also give rise to the same second derivative term with 
a redefined value of  $\mu$. 

The term $u(x)$ describes a possible spatial relative bias which in
general can be made time-dependent.
To quantify the description of motion,
we define a function of space-time $u({\bf r},t)$ which we call 
potential frequency and we write the total frequency operator 
as two contributions. In momentum space basis and in three-dimensions,
the frequency operator can be written as
\begin{eqnarray}
\hat \omega (\hat {\bf k}) = {1 \over {2 \mu}} \hat {\bf k}^2 + 
 u({\bf r},t).
\end{eqnarray}
Here we would like to identify the frequency and wave-vector with the 
energy and momentum of a particular mode; they are different words for the 
same observable in macroscopic mechanics. To make contact with 
experimental results we need to use the same notions and units for these 
quantities. This implies that when we use units such that 
 $\hat H = \hbar \hat \omega$, 
 and  $\hat {\bf p} = \hbar \hat {\bf k}$,  
the energy and the momentum operators are the same as the frequency and 
wave-vector operators.   
Note the absence of Planck's constant from these relations and from
Eqs.~(\ref{freqoperator},\ref{frequncertainty},\ref{waveoperator},\ref{waveuncertainty}).
Planck's constant enters in Quantum Mechanics because of the traditional 
or historical way of evolution of the description of nature starting 
from Newtonian notion of {\it mass} which in our notation is $m = \hbar\mu$. 

With these identifications, Eq.~\ref{schr-equation} is 
the Schr\"odinger equation with 
\begin{eqnarray}
\hat H & = & \hbar \hat \omega = {{\hat p^2} \over {2 m}} + \hat V({\bf r},t),
\label{hamilton}\\
\hat {\bf p} & = & \hbar \hat {\bf k}, \hskip 0.2 in 
\hat V({\bf r},t)= \hbar \hat u({\bf r},t), \hskip 0.2 in m = \hbar \mu.
\end{eqnarray}
Schr\"odinger's takes the form of Newton's
equation in the limit where, in any given eigenstate, the contribution of 
the first term in Eq.~\ref{hamilton} is much smaller than
the contribution of the second term. 
In addition, classical mechanics describes the
behavior of an ensemble of a huge number $N$ of such indivisible microscopic 
systems together. Therefore, the notion of the energy used in classical
mechanics is $\sum_{i=1}^N \hbar \hat \omega_i$ and the momentum of the system
will be $\sum_{i=1}^N \hat {\bf p}$.

\subsection{Relativistic Quantum Mechanics}
The validity of relativistic mechanics is additional support for the 
starting point of the present paper that everything that happens,
takes place in consciousness. Namely, space and time are not independently
existing notions but they are related through the fact that 
periodic variation (frequency) and spatial variation (wave-vector)
are both changes related to events in consciousness. 
In units where we measure distance 
by the time it takes light to ``travel'' it, frequency and wave-number 
for light are related through the relation $\omega = c k$.
This oneness of space and time is ultimately linked to the oneness
of change (namely motion is change) as the only generalized event that can 
possibly occur in consciousness.

In the case of relativistic mechanics, where we need to impose 
invariance under Lorentz transformations, using the four-vector
notation $c k^{\mu} = (\omega,c {\bf k})$, where $c$ is the speed of light, 
we have 
\begin{eqnarray}
k^{\mu}k_{\mu}= \omega^2-c^2 k^2 = \Bigl (mc^2/\hbar\Bigr )^2,
\label{lorentz}
\end{eqnarray}
and by substituting the operators given by Eq.~\ref{waveoperator} and
Eq.~\ref{freqoperator}   for $k_i$ and 
$\omega$, respectively, the Klein-Gordon equation is obtained
\begin{eqnarray}
\Bigl [-\nabla^2+{{1} \over {c^2}} \partial^2_t + {{c^2 m^2} 
\over {\hbar^2}}\Bigr ]
| \psi \rangle = 0.
\label{Klein-Gordon}
\end{eqnarray}
In the case of the dispersion relation $\omega^2(k)=c^2 k^2$
the wave equation for a massless particle is obtained.

Similarly, while on the one hand we impose invariance under Lorentz 
transformations, we may assume that the existence of a vector, such as the spin, 
breaks the rotational symmetry of space. Following Dirac, 
the following frequency operator is obtained by taking 
the square root of the operator in Eq.~\ref{Klein-Gordon}:
\begin{eqnarray}
\hat \omega({\bf k}) = c {\vec \alpha} \cdot {\hat {\bf k}} + 
 m c^2 {\hat \beta},
\label{dirac}
\end{eqnarray}
where ${\vec \alpha} = 
(\hat \alpha_1,\hat \alpha_2, \hat \alpha_3)$ and $\hat \beta$ are four 
Hermitian operators acting on the spin variables alone. The squares of 
these operators
are unity and their components anticommute, in order for the equation 
$\hat \omega^2  |\psi \rangle = - {\partial_t}^2
|\psi \rangle$ to be the same as the Klein-Gordon equation. Namely,
\begin{eqnarray}
 ( \hat \alpha_i)^2 & = & 1, \hskip 0.3 in 
\hat \alpha_i \hat  \alpha_j + \hat  \alpha_j \hat  \alpha_i=0 
\hskip 0.1 in (i\ne j),\\
\hat \beta^2 & = & 1, \hskip 0.2 in 
\hat  \alpha_i \hat  \beta + \hat \beta \hat \alpha_i=0.
\end{eqnarray}
By substituting the frequency (Eq.~\ref{freqoperator}) and 
wave-number (Eq.~\ref{waveoperator})
operators in Eq.~\ref{dirac}, the Dirac equation is obtained.

\section{\large MEASUREMENT AND STATE VECTOR COLLAPSE}
\label{measurement}

Theory does not describe what actually happens {\it independently} of 
the operation of consciousness, it describes what is observable in
consciousness.  It elegantly describes
 our experiences as ``events'' in consciousness caused 
by consciousness' operation. 
This implies that the theory should describe at once
the process of observation {\it together} with 
the description of the event, as opposed to describing separately what happens
and then leaving  the description of the process of observation for 
a later stage.  Namely, as it has been repeatedly discussed, 
we cannot possibly describe what happens outside consciousness,  
without introducing 
the presence of the operational observer in the very description 
of what happens. What instead theory describes is the very process 
of operational observation. This implies that in order to understand 
the process of measurement we do not need an additional theory of measurement, 
the theory itself  {\it is} the theory of observation. 

Theory should describe what we operationally do to observe. 
It describes how consciousness operates upon itself
in an event of observation and what the potentialities of observation are 
depending on what this operational process does.
Therefore,
there is no difference between the
theory of what ``happens'' in nature and the theory of measurement. 
Quantum theory is the description of what happens in consciousness,
namely what happens during the process of observation
and how we should describe the evolution of the state of
potential consciousness between observations. 

While there is only one consciousness,  particular observing 
instruments related to consciousness observation sites can reflect 
particular events. 
A particular instrument or observation site of consciousness 
is realized when the totality is divided into an observing instrument
and the rest which plays the role of the observed.
A particular measurement consists
of a ``question'' that consciousness has decided to ``ask'' by a) 
sectioning the whole into an observed and an  observing instrument. 
The way this division is chosen by consciousness reflects the nature of the
question to be asked. The experimental instrument is used only 
to materialize, or to operationally apply, or to reflect
the question. This instrument is made to mimic the  
operation of consciousness known from experience (see discussion of 
Sec.~\ref{brains} and Sec.~\ref{mathematical}). b) The question 
is operated by allowing the reunion (interaction) between the observing 
instrument and the observed and this reunion forces a changed state of 
potential consciousness c) this changed state is measured relative
to the state of potential consciousness prior to the application 
of the question. This process causes an experience or event in consciousness
and in the case of quantum theory it corresponds to the state collapse.

Let us divide the universe, the closed system, into an observed 
subsystem S and an observing instrument O.  
Let the eigenstates of S be
denoted by $|\alpha\rangle$ and the eigenstates of the observing instrument 
be $|a )$. These eigenstates among other quantum numbers
are characterized by a quantum number $a$ which are the
eigenvalues of an observable $\hat a$ which is a property of O. 
For the system O to play the role of a measuring device of the
property $\hat \alpha$ of S the following are required:

1) The eigenvalues of $\hat a$ must be 
in one to one correspondence with the eigenvalues 
$\alpha$ of an observable $\hat \alpha$  of S. This correspondence
is declared by the function $a=f(\alpha)$ and it can be used as
the measuring scale. Namely, by observing the value of
$a$ on the measuring scale of the instrument O, we actually observe 
the corresponding value $\alpha$ characterizing S. 

2) While in general the state of the combined system
after their interaction is
\begin{eqnarray}
| \Psi \rangle = \sum_{\alpha, a} \Psi_{\alpha,a} | \alpha \rangle 
| a ),
\end{eqnarray}
for O to play the role of an observing instrument, the choice of the
system O, (namely, the particular separation of the whole into an observing
instrument and the observed) should be restricted in
such a way that  the state of the whole system after their mutual interaction
is the following linear combination\cite{von-neumann,london}
\begin{eqnarray}
| \Psi \rangle = \sum_{\alpha} \Psi_{\alpha} | \alpha \rangle 
| f(\alpha) ).
\end{eqnarray}
Namely, in the linear combination, a pair of states 
$| \alpha \rangle | a )$  has a non-zero contribution only if
$a$ and $\alpha$ are eigenvalues of the instrument's observable 
$\hat a$  and  of S's observable $\hat \alpha$, such that $a=f(\alpha)$.
If we have made a table of these corresponding states by ``looking'' at 
the state of  the system O, we know in which state the system S is after 
the measurement.

In our case, the whole system includes the body of the observer. 
In particular,
the measuring instrument could be any part of the body of the observer.
However, even in this correlated state of instrument and object, there is 
no possibility of a collapse. What would cause the combined system
to choose a particular state, i.e., a particular combined state
$| \alpha \rangle  | a=f(\alpha) \rangle$ ? What would make the 
whole system decohere? There is nothing outside of it to help it decohere
itself.

Consciousness is the only agency which can make that choice. 
This, from moment to moment, different experience of the universe, 
is what causes the collapse and this point has been appreciated some 
time ago (see, e.g., \cite{london}). 
To understand  the wave function collapse in terms of very simple 
examples of consciousness' operation at the so-called personal subjective
world, the reader is referred to Secs.~\ref{operation}, \ref{matter} and 
\ref{brains}.
There, it was discussed that the state of potentiality,
 when observed, becomes a definite state representing a definite
conscious quality and therefore it acquires a definite value.  

Hence, within this theory, there is no puzzle in the so-called 
Schr\"odinger's cat paradox\cite{cat}  because of the Universality
or non-locality of consciousness;
as discussed, there is no separation between what happens and what
is measured; namely, there is no issue with the state vector collapse
and if an event happens, through the projective action of consciousness, 
it has occurred in one and the same consciousness.

The following discussion is related to the  Einstein, Podolsky and 
Rosen (EPR) paradox\cite{EPR}
and the results of the experiments by Aspect el al.\cite{aspect} as well
as the attempts recently made for various forms of 
teleportation\cite{teleportation}.
Starting from the character of consciousness we have shown that
space and time are observables and they do not exist ``out there'' 
independently of 
events in consciousness (Universal and particular). 
While events happen in consciousness,
the events themselves can be characterized by space-time labels only 
when these events and the measurement of space time coordinates 
can be simultaneously observed (and their observation is not incompatible 
with the observation of their complementary observables) by the 
observing instrument. Therefore, because of the non-locality of consciousness, 
an observation
which is caused by some action at a particular position in space,
influences the entire universe. In the particular example of Bohm's\cite{bohm}
formulation of the EPR paradox when the observation occurs
anywhere, the pair of spins {\it together as a single event} is born in 
consciousness from the state of potential consciousness, 
namely from the spin-singlet state 
describing a pair of correlated spins. In other words, in this case
it is not possible to observe just one spin; the very observation of
that one spin is, at the same time, observation of the other spin. 
Therefore, because of the non-locality 
of consciousness there is no paradox.
 
Causality, on the other hand, applies to two different or 
separate events in consciousness which are both characterized
by definite space and time labels as observed. 
For example, if an event occurs, where a measurement 
is made of the position 
of the particle  at a measured instant of time $t_1$ and then in a separate
event its position is measured at a different measured instant of 
time $t_2$, the second event should lie in the light cone 
which has the first event as its origin. 
In addition, the evolution of the states of potential consciousness 
between such operations of consciousness is deterministic,
bound by law and governed by cause and effect.

In any attempt to understand the process of any felt experience, namely
how it occurs using a mechanistic theory, even using quantum mechanics,
the subject or consciousness itself will never be ``seen''\cite{note2}. 
Therefore, the result of a particular operation of 
consciousness cannot be predicted, only the statistical result of
many such measurements is predictable. Quantum theory allows for
this intervention of consciousness, namely via the projection process
where the result of this projection is observed as a
destructive interference.  A good example is equilibrium
quantum statistical mechanics, where in the canonical ensemble for example,
one assumes that there is such a phase decoherence which allows one
to consider the {\it trace} of $\hat \rho \hat O$ ($\hat \rho$ is
the density matrix and $\hat O$ an operator representing an observable)
as an observable only. This is so because of phase decoherence
 introduced by the multiple interaction 
of the bath, with which the system is in equilibrium. This interaction
is actually nothing but an external (to the subsystem) ``measuring'' 
instrument through which the
subsystem is continuously observed or projected. Through such
an overwhelmingly large number of observations of an observed 
subsystem which are all averaged out, we are allowed to introduce 
the notion of  temperature and entropy of the subsystem. 

In addition, non causal evolution can exist because of consciousness' 
choice of the dividing line, which bisects the whole into an observed and 
observing instrument. Therefore, consciousness is the ultimate judge
that simply makes the choices about what questions to ask. Through
such choices the universe evolves in a direction prepared by 
the sequence of all these events in consciousness.
This process requires the division of the observed universe into an
observed part and into an observing instrument. Consciousness participates 
in this division silently through the choices and the process of 
various projections made {\it coherently} on the various material 
parts of the instrument, so that the action as a {\it whole} 
leads to a coherent measuring instrument
made for the particular reason of measurement (or reflection) and 
requires no external energy and no external material action.
Notice that this choice, which is the moment of the wave-function collapse, 
costs no external energy at all.

\section{\large SUMMARY}

A theory of consciousness was presented from which  quantum theory follows
as the quantitative description of the operation of consciousness on a
state of potential consciousness. The so-called material aspects of nature
are experienced due to events in our particular sub-streams of a Universal 
stream of conscious flow. These events give rise to conscious qualities,
such as concepts, sensations, feelings, through which we experience
the world. In addition,  through the conscious process of 
objectivation\cite{note2} which is part of
the general perception process, they are projected into 
``actual'' events.  The persistence
of the material Universe ``when no one is looking'' is due to our postulate
that our streams of
consciousness are sub-streams of a Universal conscious flow.
When an event occurs, it happens in the Universal consciousness
directly or through the particular sub-streams. Notice that we can 
describe everything using exclusively aspects of a Universal consciousness.

The so-called seat of consciousness is not to be 
found anywhere in particular because the objects are products of
consciousness; instead what can be perceived directly is not 
consciousness, but rather
the process of the operation of consciousness and the events 
which occur in consciousness. In order to discuss the operation of 
consciousness, we introduce the state of {\it potential or unmanifested 
consciousness} as a 
state representing the contents (or constructs, or abstractions, or ideals,
which manifest themselves as conscious qualities or qualia, when they
become experiences)  of all previous experiences each assigned 
a weight to be related to its probability of its projection when
a future experience takes place. 
Consciousness operates on this state of potentiality and there are
two different operations;
the creative operation of consciousness and the operation of
conscious inquiry. In addition, the potential consciousness is also
a tool to describe intuition\cite{aristotleintuition,spinoza,jung0,whitehead}
or creative advance\cite{whitehead}, a state of ``pregnancy'' of
consciousness and a tool to describe the process of evolution.

The creative action of consciousness can be thought of as 
an operation of an idea or a concept (which strictly speaking has no exact 
material representation) on the state of potential 
consciousness. This causes a change of the state of potential
consciousness. 
 Consciousness either perceives a change in its own state, which is 
recognized by the process of measurement or projection or
thought, or asks a question (inquiry) by acting on the potential state. The
result of such an operation is to manifest the conscious concept either
(a) on the measuring instrument or (b) in the case of the brain, 
the action of the 
conscious-concept is to actualize the neural correlate in the brain,
which is identified with the so-called collapse of the quantum state.
When an event happens, it always happens due to the action of the
conscious concept from the Universal stream or sub-streams of 
consciousness on the state of potential consciousness.

A simple framework to quantify the description and to apply it to 
the science of perception and more generally to the science of
consciousness as well
as physics, is to work with operators acting in linear spaces representing
potential states of consciousness which are based on a few
basic concepts.
An operator represents the direct operation of the conscious-concept 
and, in general consciousness, on the
state of potential consciousness. The outcome of measurement 
or experience is (i) to collapse the state describing the potential 
for consciousness to a particular state representing a corresponding
conscious concept and (ii) the value of the observable is represented as
the result of evaluating the overlap between the change
of the state of potential consciousness after the action of
consciousness and its own state of potentiality before the action of
the operator; namely, consciousness uses the previous
state of potential consciousness as the measure to evaluate  the
quality and degree of change.

We begin from elementary contents of the Universal and particular
streams of consciousness: (a) the notion of frequency and periodic
change and its complementary concept of time, (b) the notion of
space and its complementary concept of motion in space. We use 
neuro-physiological evidence to argue that
consciousness can directly observe frequency and wavelength, independently
of the experience of time and space. We also find that the equation of
motion of the state of potential consciousness, when it is restricted
to the potential observation of the spatial position of a particle,
is Schr\"odinger's equation, where the state of potential consciousness is
identified with the wave-function of the particle. 

Furthermore, we show that this theory is free from the paradoxes and 
puzzles present
in the usual interpretations of quantum theory, such as the
EPR paradox and the Schr\"odinger cat paradox. 
In addition,
the two well-known postulates of von Neumann's quantum theory 
of measurement follow from these more general philosophical
ideas. Therefore, the theory is also
free from the  well-known problem of wave-function collapse 
which appears in the quantum theory of measurement.

\section{\large ACKNOWLEDGMENTS}
This work is dedicated to the memory of my son, 
Jacob Anthony Manousakis, from which it was inspired.
I would like to thank Professors N. Antoniou, F. Flaherty, C. Ktorides,
P. Varotsos and D. V. Winkle for
comments, and S. Barton, J. Ryan, N. Sarlis and I. Winger 
for proof-reading the manuscript at an early state of its development.
In addition, I would like to thank
the three anonymous referees for their comments; because of the effort made
to respond to these comments,
the manuscript has significantly improved from the
originally submitted version.
\section{\large APPENDIX}
\label{appendix}
\subsection{Mathematical Description of the Operation of 
Consciousness: An Example}
A child's conceptual development might proceed first by conceiving 
the integer numbers, then the fractions
and then the irrational numbers. The concept of irrational numbers
is introduced operationally as a solution to algebraic equations, say, for
example $\sqrt{2}$, as the solution to the equation $x^2 = 2$.
However, an exact solution to this equation can never be a
content of a particular brain (due to its limitation) but only fractional 
approximations  to the solution  or sequences with 
limit the solution. While with this concept of ``getting arbitrarily close'' 
satisfies us most of the time, there are phenomena, such as chaos, where 
even the slightest departure from
exactness leads to qualitatively different states. Pythagoras demonstrated,
however,  that $\sqrt{2}$ can be grasped through a geometrical 
operation of consciousness if we 
abandon our attachment to one dimension (real axis) and go to two dimensions.
The reason is that while consciousness is undivided, with no beginning
and no end, a particular instrument of it (e.g. the brain), is always limited.
Therefore, such an operational definition through an infinite series of 
approximants may not always be satisfactory. 

To demonstrate our point,  let us go ahead and use such an 
iterative process which provides closer and closer approximants to the
solution of the equation $f(x)=0$. In particular let us adopt the
Newton-Raphson method as the operational definition of the solution 
of the equation $f(x)=0$, where the following recursion relation
$x_{n+1} = x_n - {{f(x_n)} / {f'(x_n)}}$ is iterated
starting from $x=x_0$. Through this process the solution is conceptualized 
by the intersection of the curve $f(x)$ and the real axis.  

Now, let us define the Newton-Raphson operator as follows:
\begin{eqnarray}
\hat O | x_n \rangle = | x_{n+1} \rangle, \hskip 0.3 in 
x_{n+1} = O(x_n) \equiv x_n - {{f(x_n)} / {f'(x_n)}}.
\label{nr-operator}
\end{eqnarray}
In the case of the equation $f(x)=x^2-2=0$, the recursion relation is
$x_{n+1} = (x_n^2+2)/2 x_n$.
If we iterate this equation for a large number of operations
we find that the limiting distribution approximates a delta function
at $x_0=\sqrt{2}$. Using the language of the operational consciousness
this can be formulated as follows. The operator $\hat O$ 
creates the potential solution of this equation defined above.  When 
the state $|x_{n+1} \rangle =
\hat O | x_n \rangle  $ and the previous state $| x_{n} \rangle$ have 
large overlap we claim that the solution is observed. 
This is how we decide that we have found the solution,
namely when $\langle x_n | \hat O | x_n \rangle = 1$,
within a resolution defined by our computer precision.
In order to make sure that we found a solution independent of the
initial condition, we may start from a state
$| \psi_0 \rangle = \sum_{l=1}^M | x^l_0 \rangle$ and after
application of the operator $\hat O$ several times and we stop when
the overlap $\langle \psi_n | \hat O | \psi_n \rangle$ is maximum
(or unity, if we normalized the states $| \psi_n \rangle$).

This operational procedure is used here to simulate the operation 
of thought and of consciousness. Namely, the operational procedure which
applies a question on the state of potential-consciousness and 
obtains an answer by comparing the states before and after the application. 
In addition, it can be also used to represent the experimental 
measurement procedure. 

Next, let us assume that we wish to find the solution of the equation
$f(x)=x^2+1 = 0$, using the same operational definition of 
what is meant by solution.  The Newton-Raphson operator for this case 
is such that $| x_{n+1} \rangle  = \hat O| x_n \rangle $ with
$x_{n+1} = (x_n^2 -1)/(2x_n)$.
Here, we assume that our consciousness has only experienced
real numbers. A solution to the equation $f(x)=0$ does not exist 
on the real axis.  
However, to demonstrate our point let us say that we cannot grasp 
other notions such as the ``imaginary'' numbers, because we do not
have such direct experience. Therefore, let us insist on looking 
for the solution on the real axis using the Newton-Raphson method.
This is meant to parallel the fact of our firm belief, acquired from our
macroscopic experience, that every ``particle'', while in 
a state of motion, is also somewhere in space. 
In addition, this is meant to parallel the fact that these two states 
are incompatible states of consciousness; namely when the solution is
an imaginary number it cannot be placed on the real axis at the same time.
Therefore the Newton-Raphson algorithm, is used here as an 
``experimental'' device to materialize, or operationally apply, or reflect
the question, where is the solution to the equation $x^2=-1$ 
on the real axis? The incompatibility is that the state 
$| i \equiv \sqrt{-1}\rangle $ has nothing to do with real numbers.
We will find out using our definition of what we mean
operationally by a solution to an equation (which also is meant to 
map our ``experimental'' operations), that the question,
``where is the solution?'', is not a good question. A better
question would be, ``what is the potential solution?''. 
These statements will become 
clear through this example. 

If we iterate for a very long time, we will notice that the method \
passes many times 
arbitrarily close to any real number. However, the neighborhood 
of certain numbers are visited more frequently that others.
The probability density of visiting a particular small region
near a number $x$ is plotted in Fig.~\ref{distribution} as found
by iterating the equation $x_{n+1} = (x_n^2-1)/2 x_n$ about 200,000 times.
Can we give any practical meaning  or interpretation to this 
distribution?
We have already paralleled the computerized search for the solution to 
this equation to the operation of consciousness or an experimental 
procedure to determine the position of a particle.
Let us further assume that this procedure requires a very large number 
of iterations (such as of the order of the
Avogadro number) because each operation of projection (or measurement) 
in practice is carried out 
by a macroscopically  large number of microscopic  processes. 
Another reason is that since we are accepting as a solution the converged
value of this process, we have to carry out a large
number of such iterations.
If it is applied   $10^{23}$ times we will obtain one 
value of $x$. If  we then repeat the procedure by using the same number of
iterations, plus one more, then two more and so on,
we can obtain different outputs, but they 
belong to the distribution given in Fig.~\ref{distribution}.
Therefore in this case we might say that the experimentally determined
value of this variable is random  but it has a definite probability 
distribution given by that of Fig.~\ref{distribution}.

The operator $\hat O$ has no unique inverse because the 
two operators $\hat O^{-1}_{\pm}| x \rangle  =  | x \pm \sqrt{x^2+1} \rangle$
have the property of the inverse operation, namely
$\hat O \hat O^{-1}_{\pm} = \hat 1$. After a large number of iterations 
$n$ starting from  the  state $x_0$, the state
$\hat O^n | x_0 \rangle = | x_n = O(O(...O(x_0)..)) \rangle$
is not a well-defined output of the procedure because it depends on 
$n$. However, there is something else which is well-defined as 
a limit for $n> n_0$, where $n_0 \to \infty$.  Let us define the operator
\begin{eqnarray}
\hat P_{n_0 \to n} \equiv \sum_{m=n_0}^n\hat O^m,
\label{project}
\end{eqnarray}
where $n_0, n$ are numbers of the order of the Avogadro number and 
$n>n_0$. When this operator acts on a starting state $| x \rangle$ the 
outcome is a distribution 
\begin{eqnarray}
|\psi \rangle = \hat P_{n_0 \to n} | x \rangle
= \sum_{m=n_0}^n | x_{m-n_0} = O(O(...O(x)...)) \rangle,
\end{eqnarray}
and in the last equation, the function $O(x)$ has been applied $m-n_0$ times
on the value $x$. 
The important point is that the state $| \psi \rangle$,
in the limit $n_0 \to \infty$ $n\to \infty$ with $n=n_0+m$ and $m>>1$, 
depends {\it only on the starting value of $x$}. As we will show in the 
following subsections, the dependence on the value of $x$ is an overall 
prefactor, otherwise 
the normalized state $|\psi \rangle$ is independent of $x$.

It is useful to consider the  eigenstates of the Newton-Raphson 
operator $\hat O$ that correspond to the  eigenvalue $\lambda_{\nu}$:
\begin{eqnarray}
\hat O | {\nu} \rangle = \lambda_{\nu} | {\nu} \rangle.
\label{eigen}
\end{eqnarray}
Let us consider the equation $O(O...(O(x))...) = x$ where the function
$O(x)$ has been applied $n$ times. Given  a solution $x=x_0$ to this 
equation, the following state is an eigenstate of the operator $\hat O$:
\begin{eqnarray}
| n, x_0 \rangle = { 1 \over {\sqrt{n}}} \sum_{m=1}^{n-1} 
\hat O^m | x_0 \rangle,
\end{eqnarray} 
with corresponding eigenvalue unity.
An example of such an eigenstate is $|2,1/\sqrt{3} \rangle = 1/\sqrt{2}
(| 1 /\sqrt{3} \rangle + | - 1/\sqrt{3} \rangle)$, because $O(1/\sqrt{3})
=-1/\sqrt{3}$ and $O(-1/\sqrt{3}) =1/\sqrt{3}$.

Fig.~\ref{evolution}(a) shows the evolution of the variable $x$ 
under the action of the Newton-Raphson operator $\hat O$ starting from the
initial value $x_0=1/\sqrt{3}$ for 100 iterations. Notice the cycle 
$1/\sqrt{3} \to - 1/\sqrt{3} \to  1/\sqrt{3}$ is followed for a few 
iterations and after that the cycle is broken. The reason is the
following. A computer cannot exactly represent the irrational
number $1/\sqrt{3}$ because it only has a limited precision.
Therefore it starts from the neighborhood of that number (the truncation
of the number, say up to ten significant digit accuracy)
but a few iterations later the  value
of $x_n = \hat O^n x_0$ is very far from the original solution
corresponding to the cycle of period 2, because of the propagated error.
In fact after a few iterations the value of the $x_n$ is closer to a member
belonging to a different cycle of period $m$.
If the period $m$ is a large integer, because of the limited precision
before the end of even one cycle, we get closer to another cycle and so on.
Figs.~\ref{evolution}(b-c) give the
evolution of the variable $x$ within 1000 and 10000  iterations.
The distribution of $x$ after 200000 iterations is shown in 
Fig.~\ref{distribution} where it is compared with the 
Lorentzian (which is the exact solution, see Eq.~\ref{lorenzian}).

Eigenstates corresponding to an infinite cycle (continuum) are
of the following form
\begin{eqnarray}
\psi_{\nu}(y)  = {{\phi_{\nu}(O(y))} \over {y^2 + 1}},
\hskip 0.2 in \phi_{\nu}(y)  = \lambda_{\nu}  \phi_{\nu}(O(y)).
\label{lorenzian}
\end{eqnarray}
An obvious solution to the second equation is the constant
which implies  $\psi_{1}(y) =  1  / {\pi}(y^2 + 1)$
with eigenvalue $\lambda=1$.
This is shown by the solid line in Fig.~\ref{distribution}
which fits very well the results of the numerically implemented
Newton-Raphson method. The other solutions of the above equation
are not going to be discussed because this goes beyond the goal of this
work. 

In the limit of $n_0 \to \infty$ and $n-n_0 \to \infty$ the 
projection operator takes the form 
\begin{eqnarray} 
\hat P_{n_0 \to n} | x \rangle = c \sum_{\mu: \lambda_{\mu} =1 }  
\langle \mu | x \rangle | \mu \rangle \nonumber \hskip 0.2 in c =  n-n_0.
\end{eqnarray}

The question is why is the probability density to observe a definite value of
$x$, $|\psi(x)|^2$ and not $\psi(x)$? 
In order to measure the probability to find the value $x$ we need
to start the projection process at $x$ and measure it at $x$ or another
place $x'$.  The ``particle'' is not on the real axis, we just believe
it must be. We also know that it can be on the real axis in the sense
defined by the operational definition or by our way of 
searching.
There is no ``particle'' anywhere on the real axis unless we start the 
operational procedure which defines it, in quantum mechanics this means the
experimental procedure. Therefore, particles do not exist on our 
{\it perception screen} on their own, the act of observation creates them.  
We need to start the measuring or projection process from some
value of $x$ and ask the question what is the probability to observe
the ``shadow'' of the particle at $x$. Our procedure implies that we will
measure the ratio of the number of occurrences of the particle at $x$ 
(or between $x$ and $x+dx$)  to the total number of measurements $n-n_0$. 
This implies that the probability is proportional to
$P(x) \propto  \langle x | \hat P_{n_0\to n} | x \rangle$.
In the limit of large $n$ we find
\begin{eqnarray}
P(x) & \propto &  \sum_{\nu: \lambda_{\nu}=1}| \psi_{\nu}(x)|^2,
\end{eqnarray} 
and this agrees with what we know from the quantum mechanical 
measurement process.

\subsection{``Quantum'' Interference}

Let us see how far we can push this analogy with quantum mechanics.
One of the most important aspects of quantum theory is the so called
interference.  The two-slit thought experiment is the best known
formulation of the problem.  
Here we will consider the case of the equation $f(x)=0$ with
$ f(x)=(x^2+\delta) ((x-3)^2+\delta)$.
While there are two independent sets 
of solution one with real part equal to zero, as before, and 
another with real part equal to 3, 
our ``measurement'' (or projection) process gives interference.

\begin{figure}[ht] 
\begin{center}
\includegraphics[width=5. in]{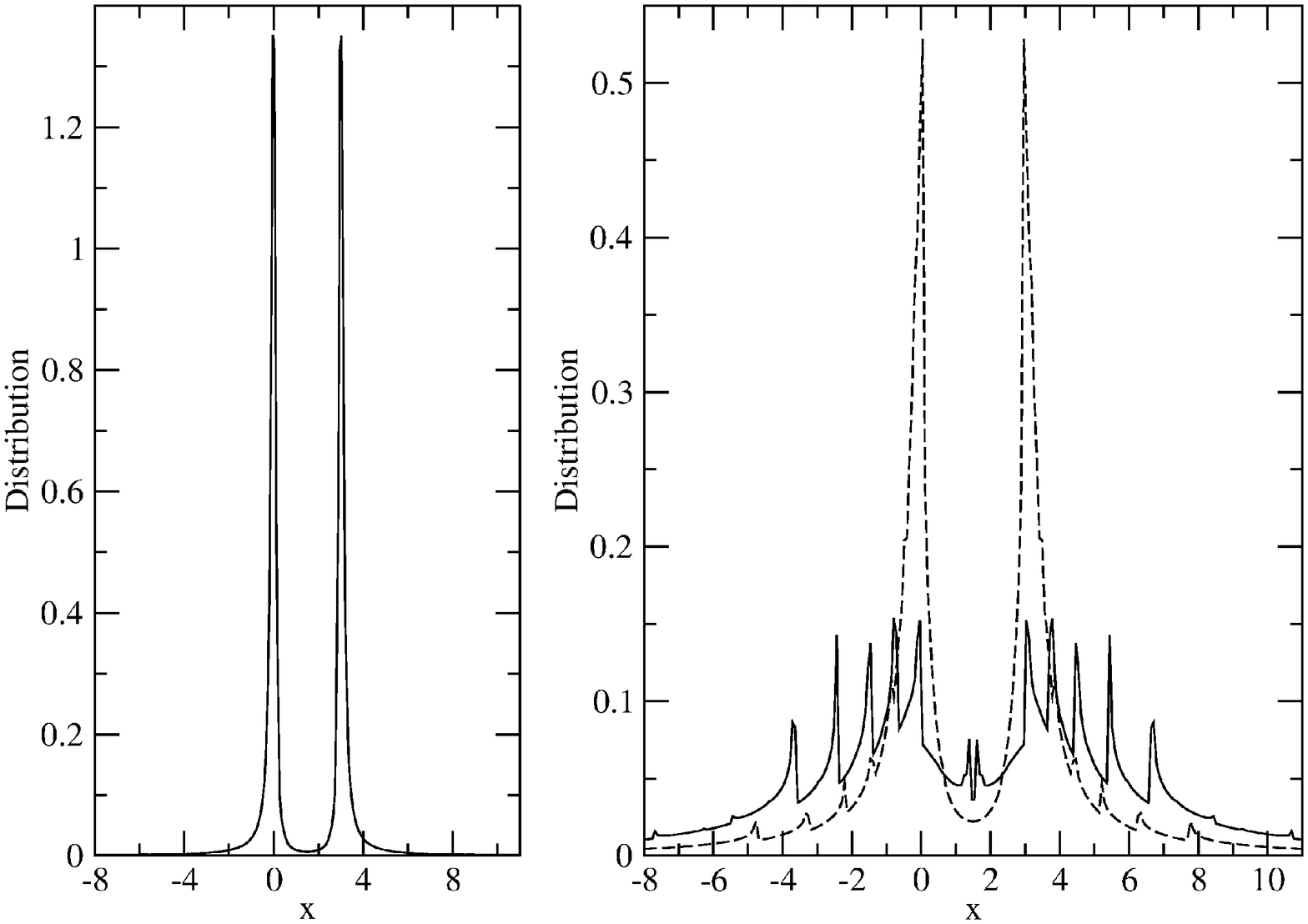}
\caption{ \label{interference}}
\end{center}
\end{figure}

First we choose a small value of $\delta=0.01$, and
when we apply the Newton-Raphson projection algorithm, we find 
the distribution shown in Fig~\ref{interference} (Left graph).
There are two Lorentzian peaks near the real parts of the solutions, namely
$x=0$ and $x=3$ and the widths of these Lorentzian distributions are of the
order of the imaginary part, namely $0.1$.  
Notice that because of the small width of these two distributions,
there is negligible overlap and the ``particle'' seems to be
either around $x=0$ or around $x=3$. 
In the right graph of Fig.~\ref{interference}, the distribution
obtained from the same projection process for $\delta=0.1$ (dashed line) and
$\delta=1$ (solid line). Notice that as the widths of the two distributions
become broader, interference peaks begin to appear. They correspond to 
values of $x$ which form cycles but they arise from bounces off both
neighborhoods, namely, the $x=0$ and the $x=3$ neighborhood.
Since the actual ``particle'' is either at $x=0\pm i \sqrt{\delta}$ or 
$x=3\pm i \sqrt{\delta}$  and nowhere on the real axis, its ``shadow'' on
the real axis, which is what we observe (because of our insistence of
asking the wrong question), appear to be in both places at once
and to interfere.

\newpage

\begin{figure}[ht] 
\begin{center}
\includegraphics[width=5.0 in]{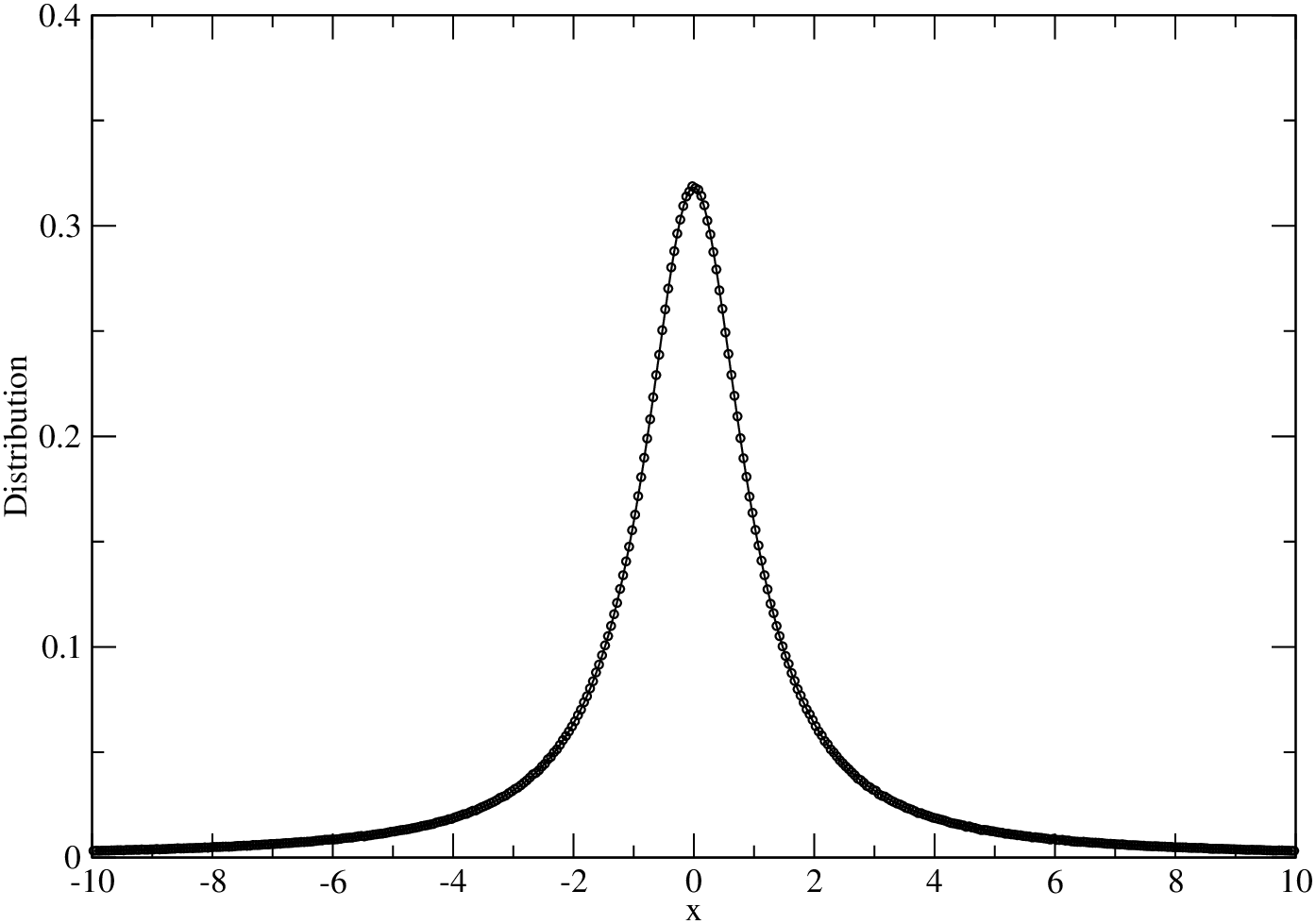}
\caption{ \label{distribution}} 
\end{center}
\end{figure}

\centerline{Figure Captions}
Fig.~\ref{distribution}:
The distribution of the steps of the Newton-Raphson iteration process
to solve the equation $x^2=-1$ within the real axis. This is also 
an eigenstate of the Newton-Raphson operator as defined 
by Eq.~\ref{nr-operator} with eigenvalue unity.

Fig.~\ref{evolution}: The evolution of the variable $x$ under 
the action of the Newton-Raphson 
operator to observe the solution of Eq.~\ref{nr-operator} starting from the
initial value $x_0=1/\sqrt{3}$.

Fig.~\ref{interference}:
Demonstration of interference. 
Left graph is obtained for $\delta=0.01$. 
Right graph:  for $\delta=0.1$ (dashed line) and $\delta=1$ (solid line).
\newpage

\newpage

\begin{figure}[ht] 
\begin{center}
\includegraphics[width=5.0 in]{fig2.eps}
\caption{\label{evolution}}
\end{center}
\end{figure}
\newpage
\newpage


\begin{thebibliography}{99}
\bibitem{wheeler} J. A. Wheeler and W. H. Zurek, {\it Quantum Theory and
Measurement}, (Princeton University Press, Princeton, 1983).
\bibitem{EPR} A. Einstein, B. Podolsky and N. Rosen,
{\it Phys. Rev.} {\bf 47}, 777 (1935).
\bibitem{cat} E. Schr\"odinger, {\it Proc. Cambridge Phil. Soc.} {\bf 31}, 
555 (1935); $ibid$ {\bf 32}, 446 (1936).
\bibitem{vandermerwe}F. Selleri, and A. van der Merwe,
{\it Quantum Paradoxes and Physical Reality}
(Kluwer Academic, Dordrecht, 1990). See also references therein.
\bibitem{bohr} N. Bohr, {\it Atomic Theory and the Description of Nature}
(Cambridge University Press, Cambridge, 1934). 
{\it Atomic Theory and Human Knowledge} (Wiley, New York, 1958).
\bibitem{heisenberg} W. Heisenberg, {\it The Physical Principles of the 
Quantum Theory} (Dover, New York, 1930). {\it Physics and Philosophy}, 
(Harper and Row, NY, 1958).
\bibitem{bohm1} D. Bohm, {\it Phys. Rev.} {\bf 85}, 166 (1952).
$ibid$, {\bf 85}, 180 (1952).
\bibitem{everett} H. Everett III, {\it Rev. Mod. Phys.} {\bf 29}, 463 (1957).
\bibitem{ballentine}L. E. Ballentine, {\it Rev. Mod. Phys.} {\bf 42}, 
358 (1970).
\bibitem{bell}J.S. Bell, and A. Aspect,  {\it Speakable and unspeakable in 
quantum mechanics: Collected papers on quantum philosophy} 	
(Cambridge University Press, Cambridge, 1987). 
\bibitem{vandermerwe2} A. van der Merwe, 
F. Selleri and G. Tarozzi, {\it Microphysical Reality and Quantum Formalism}, 
 Eds., Vols I and II (Kluwer Academic, Dordrecht, 1988).
\bibitem{stapp} H. P. Stapp, {\it Mind, Matter and Quantum Mechanics}
(Springer-Verlag, Berlin, 2003).
H. P. Stapp, {\it Found. Phys.} {\bf 10}, 767 (1980).
\bibitem{stapp2}J. M. Schwartz, H. P. Stapp and M. Beauregard, 
{\it Phil. Tran. Royal Soc.} {\bf B 360} (1458), 1306 (2005).
\bibitem{schrodinger} E. Schr\"odinger, {\it What is life? 
and Mind and Matter} (Cambridge University Press, Cambridge, 1967). 
Studying the entire book is strongly recommended and in particular
Chap. 3, pg. 126 and Chap. 4, pg. 139.
\bibitem{schrodinger2}E. Schr\"odinger, {\it Nature and the Greeks}, 
(Cambridge University Press, Cambridge, 1954).
\bibitem{wigner} E. P. Wigner, in {\it Quantum Theory and Measurement}
J. A. Wheeler and W. H. Zurek eds.,
 pg.260 and pg. 325 (Princeton University Press, Princeton, 1983).
\bibitem{von-neumann} J. Von Neumann, {\it Mathematical Foundations of Quantum 
Mechanics},  Chap. VI, pg. 417 (Princeton University Press, Princeton, 1955).
\bibitem{london} F. London and E. Bauer, in {\it Quantum Theory and
 Measurement}, J. A. Wheeler and W. H. Zurek eds.,
 pg. 217 (Princeton University Press, Princeton, 1983).
\bibitem{pauli} W. Pauli and C. G. Jung, {\it Atom and the Archetype,
Pauli/Jung, Letters, 1932-1958}, ed. C. A. Meier, 
(Princeton University Press, Princeton, 2001).
\bibitem{penrose} R. Penrose, {\it The Emperor's new Mind} (Oxford University 
Press, New York, 1989), and
{\it The Shadows of the Mind}  (Oxford University Press, New York, 1994).
\bibitem{nanopoulos}N. E. Mavromatos and D. V. Nanopoulos, {\it Int J. Mod. 
Phys.} {\bf B 12}  517(1998).
\bibitem{poppel} E. P\"oppel, {\it Trends Cognit. Sci.} {\bf 1}, 56-61 (1997).
\bibitem{aristotleintuition}Aristotle, {\it Posterior Analytics} 
II 19, 99b28-29:
 \greektext
`` e\'i d\'e lambanomen m\'h \'eqontes pr\'oteron, p\'ws \'an gnwr\'izoimen 
ka\'i manj\'anoimen ek m\'h pro\"uparqo\'ushs gn\'wsews.''
\latintext
\bibitem{kant} I. Kant, {\it Critique of Pure Reason} (1781), see translation
in English by P. Max M\"uller (Anchor books, New York, 1966).
\bibitem{spinoza} B. Spinoza, {\it Ethics}; 
edited and translated by G.H.R. Parkinson  
(Oxford University Press, New York, 2000).
\bibitem{whitehead}A. N. Whitehead, {\it Adventures of ideas}, 
pg. 228 (Macmillan, New York, 1933).
\bibitem{jung0}C. G. Jung, {\it Psychological Types}, pg. 567
(Princeton University Press, Princeton, 1971);
C. G. Jung, {\it The integration of the personality},
(Farrar and Rinehart, New York, 1939);
C. G. Jung, {\it Psychology of the unconscious},
(Dodd, New York, 1916).
\bibitem{potentiality} The idea of something ``potentially existing'' 
was discussed by Aristotle, see, e.g.,  {\it Physics}, 186a1-3.
 \greektext
%`\char0185sper o\char0206k \char0226ndeq\'omenon ta\char0206t\'on 
%\char0227n te ka\'i poll\'a  e\char0218nai,
% ..., m\'h t\char0130ntike\'imena d\'e, 
``..., \char0234sti g\'ar t\'o \char0227n ka\'i dun\'amei ka\'i 
\char0226nteleqe\'i\d{a}.'' 
\latintext
This can be translated as follows:
``..., because the one exists in {\bf potentia} and in {\bf actuality}.''
\bibitem{parmenides} Parmenides, {\it On Nature}, Pre-Socratic Greek 
Philosopher,
born in 510 B.C.  See {\it The fragments of Parmenides}, A. H. Coxon,  
(Assen, Netherlands, 1986). See also 
Ref.~\cite{parmenides2}
\bibitem{parmenides2} {\it Parmenides}, Presented by Plato, pg. 920, 
 Ref.~\cite{plato}.
\bibitem{plato} Plato, {\it Collected Dialogs}, Eds. E. Hamilton and 
H. Cairns (Princeton University Press, Princeton, 1980).
\bibitem{vedanta} S. Vivekananda, {\it The 
Complete Works of Swami Vivekananda}, Mayavati memorial edn.
(Advaita Ashrama, Calcutta, 1965).
\bibitem{krishnamurti} J. Krishnamurti and D. Bohm, {\it The Ending of Time}
(Gollancz, London, 1985).
\bibitem{note5}We postulate that the seat of consciousness cannot
be matter-energy itself because matter is itself an experience of 
consciousness; namely the experience of matter is given us 
{\it posteriori} but {\it  that} which perceives matter, that which has
the experience, must be ready for the experience to 
occur {\it a priori}\cite{kant}.
\bibitem{lonergan}E. Webb, {\it Philosophers of Consciousness},
Chapter 2, ``B. Lonergan, Consciousness as experience and operation'',
pg. 53 (University of Washington Press, Seattle, 1988).
\bibitem{note1} The original meaning of the Greek word ``phenomenon'' 
is ``appearance'', namely, that which appears in consciousness.
\bibitem{note2} When a sentient being is examined to study ``his'' 
consciousness using all presently available instrumentation, the being
is turned into an object (See Ref.~\cite{schrodinger}, Chapter 3, 
``The principle of objectivation''). Subject is the experience of oneself. 
For example, if we follow the nerve excitation caused by the molecules
of a flower which interact with those of his nose we will never ``see''
or experience the aroma. All we will be able to see is the electromagnetic 
imprint, the pointer which ultimately the subject experiences. 
Some people are inclined
to think that this is not the final stage, that somehow another part of
the brain has looked at this imprint and interpreted it. However, we 
have already included this, namely, the imprint we are considering is 
the one produced in the brain after this process,
namely, it is the collective neural excitations including the
neurons that process all the series of signals and their translations
to other signals. See also, Ref.~\cite{schrodinger}, Chapter 6, ``The 
mystery of the sensual qualities''.
\bibitem{hubel} D. H. Hubel,  and T. N. Wiesel. {\it J. Physiol.} {\bf 148},
574-591 (1959).
\bibitem{vision} D. H. Hubel, {\it Eye, Brain, and Vision} (Scientific American
Library Series, New York, 1995).
\bibitem{eccles} J. C. Eccles, {\it How the self controls its brain}
(Springer-Verlag, Berlin, 1994).
\bibitem{visionRMP} F. Rieke and D. A. Baylor, {\it Rev. Mod. Phys.}
{\bf 70}, 1027 (1998).
\bibitem{receptors} K.-W. Yau, and D. A. Baylor, {\it Ann. Rev. Neurosci.}
{\bf 12}, 289 (1989).
\bibitem{singer} A. K. Engel, P. Konig, A.K. Kreiter, T. B. 
Schillen, W. Singer, {\it  Trends Neurosci.} {\bf 15}, 218 (1992).
C. M. Gray. {\it J. Comput. Neurosci.} {\bf 1}, 11 (1994).
P. Fries, J.-H Sch\"oder, P. R. Roelfsema, W. Singer, A. K. Engerl,
{\it J. Neurosci.} {\bf 22}, 3739 (2002).
\bibitem{direction-neurons} C. L. Baker, Jr. and M. S. Cynader, 
{\it J. Neurophysiol.} {\bf 55}, No 6, 1136 (1986).
\bibitem{bohm} D. Bohm, {\it Quantum Mechanics} (Dover, New York, 1979).
\bibitem{aspect} A.  Aspect, P. Grangier and G. Roger,
{\it Phys. Rev. Lett.} {\bf 49}, 91 (1982); $ibid$, {\bf 47}, 460 (1981);
A. Aspect, J. Dalibard and G. Roger, {\it Phys. Rev. Lett.} 
{\bf 49}, 1804 (1982).
\bibitem{teleportation}D. Bouwmeester,  {\it et al.}  
{\it Nature} {\bf 390}, 575 (1997). 
D. Boschi, {\it et al.} {\it Phys. Rev. Lett}. {\bf 80}, 6, 1121-1125 (1998);
I. Marcikic, {\it et al.}, {\it Nature} {\bf 421}, 509 (2003);
M. Riebe, {\it et al.}, {\it Nature} {\bf 429}, 734(2004);
M. D. Barrett, {\it et al.}, {\it Nature} {\bf 429}, 737 (2004).

\end{thebibliography}
\end{document}